\documentclass[aps,prc,preprint,showpacs,showkeys]{revtex4}
\usepackage{amssymb}
\usepackage{amsmath}
\usepackage{graphicx}
\usepackage{bm}
\usepackage{epsf}

\begin{document}

\title{Effects of finite size and symmetry energy on the phase transition
of stellar matter at subnuclear densities}
\author{S. S. Bao}
\affiliation{School of Physics, Nankai University, Tianjin 300071, China}
\author{H. Shen}~\email{shennankai@gmail.com}
\affiliation{School of Physics, Nankai University, Tianjin 300071, China}

\begin{abstract}
We study the liquid-gas phase transition of stellar matter with the inclusion
of the finite-size effect from surface and Coulomb energies.
The equilibrium conditions for two coexisting phases are determined
by minimizing the total free energy including the surface and
Coulomb contributions, which are different from the Gibbs conditions
used in the bulk calculations. The finite-size effect can significantly
reduce the region of the liquid-gas mixed phase.
The influence of the symmetry energy on the liquid-gas phase transition
is investigated with the inclusion of finite-size effects.
It is found that the slope of the symmetry energy plays an important
role in determining the boundary and properties of the mixed phase.
\end{abstract}

\maketitle


\section{Introduction}
\label{sec:1}

The nuclear liquid-gas phase transition is of great interest
because of its importance in heavy-ion collisions and
astrophysics~\cite{PR05,LiBA08,Pal10b,Hemp10,Sagu14}.
At densities below saturation and temperatures
lower than $\sim 20$ MeV, stellar matter may be in a liquid-gas
mixed phase which plays a crucial role in various
astrophysical phenomena such as supernova explosions and neutron
star formation~\cite{Pal10b,Hemp10}.
In past decades, tremendous experimental and
theoretical efforts have been devoted to the study of the
liquid-gas phase transition in symmetric and asymmetric nuclear
matter~\cite{Jaqa84,Latt85,Sero95,PRL95,PR04,RPP05,PPNP08,Pal10a}.
The early theoretical studies of nuclear liquid-gas phase
transition~\cite{Jaqa84,Latt85} were performed by using Skyrme
interactions. A very detailed analysis of the liquid-gas phase
transition in asymmetric nuclear matter was reported by M\"{u}ller
and Serot in Ref.~\cite{Sero95}, where a relativistic mean-field
model was employed and the effect of symmetry energy was discussed.
The authors of Ref.~\cite{Sero95} argued that asymmetric nuclear matter
may present different types of spinodal instabilities:
a mechanical instability associated with fluctuations in the baryon
density (isoscalar) and a chemical instability associated with
fluctuations in the proton concentration (isovector).
In fact, it was pointed out in Ref.~\cite{Marg03} that spinodal
instabilities in asymmetric nuclear matter should not be
classified as mechanical or chemical,
but only one type of instability determined by the curvature
of the free energy. In the past decade, there have been numerous studies
on instabilities of nuclear and stellar matter by analyzing curvature
properties~\cite{PR04,Bara05,SSA06,Duco06,Duco07,Hemp13,Well14,Well15}.
In most of the investigations, properties of the liquid-gas mixed phase
were obtained from a bulk calculation, in which the phase
coexistence is governed by the Gibbs conditions and the finite-size
effects like surface and Coulomb contributions are neglected.
It was found that the inclusion of surface and Coulomb effects
has a significant impact on the critical temperature and the liquid-gas
coexistence region based on various
approximations~\cite{Jaqa84,Lee01,Pawl02,Sil04,Maru10}.
It is interesting and important to investigate the finite-size effects
on the liquid-gas phase transition of stellar matter in a consistent manner.

In this article, we study the influence of surface and Coulomb
effects on the liquid-gas phase transition of stellar matter
by using a compressible liquid-drop (CLD) model.
The matter is composed of nuclear clusters embedded in a gas of
free nucleons and electrons, where the proton charge is neutralized
by the uniform electron gas. The equilibrium conditions between the
nuclear liquid and gas phases are determined by minimization
of the total free energy
including the surface and Coulomb contributions~\cite{Latt85,Bao14b},
which are different from the Gibbs conditions derived in a bulk limit.
In previous studies using the coexisting phases (CP) method~\cite{Mene08,Bao14a},
the two coexisting phases were obtained by solving the Gibbs conditions
for phase equilibrium, and then the surface and Coulomb energies
were added perturbatively. It was shown in our previous
work~\cite{Bao14b} that the finite-size effect from surface
and Coulomb energies may be too large to be treated perturbatively
at low densities. Therefore, in the present study, we prefer to treat
the finite-size effect properly in the CLD model, where
the surface and Coulomb contributions are included not only in calculating
the properties of the mixed phase but also in deriving the equilibrium
conditions for two coexisting phases.
Recently, the authors of Ref.~\cite{Pais15} studied the pasta phase
in core-collapse supernova matter using three different approaches,
namely the CP method, the CLD model, and the Thomas--Fermi approximation.
They compared the results and found that the CLD model can give
very similar results to the self-consistent Thomas--Fermi calculation.

It is well known that the surface tension plays a crucial role
in determining properties of the liquid-gas mixed
phase~\cite{Mene08,Bao14a,Pais15,Maru05,Mene10}.
Actually, it was found in Ref.~\cite{Mene08} that a parametrized
surface tension would fail to predict the occurrence of the liquid-gas
phase transition in $\beta$-equilibrium matter.
Therefore, it is very important to determine the surface tension in
a proper way. We calculate the surface tension by using a Thomas-Fermi
approach for a one-dimensional system consisting of protons and
neutrons~\cite{Bao14a,Mene10,Agra14,Cent98,Douc00}. We consider a semi-infinite
slab with a plane interface which separates a dense liquid phase
from a dilute gas phase. With the density profiles obtained in
the Thomas-Fermi approach, we can calculate the surface tension
as described in Refs.~\cite{Bao14a,Mene10,Agra14}.

We employ the Wigner--Seitz approximation to describe inhomogeneous
stellar matter at subnuclear densities. The Wigner--Seitz cell, which consists
of protons, neutrons, and electrons, is assumed to be spherical and charge
neutral. In principle, nonspherical nuclei, known as pasta phases, may appear
in the liquid-gas coexistence region, and the geometrical structure of the mixed
phase is expected to change from droplet to rod, slab, tube, and bubble with
increase of the matter density~\cite{Bao14a,Pais15,Wata04,Wata08,Oyam07}.
In this study, we mainly focus on the influence of finite-size effects
on the boundary of the liquid-gas coexistence region. Therefore,
we consider only droplet and bubble configurations, while other pasta phases
appearing in the middle region are neglected for simplicity.
At low temperatures, heavy nuclei may form a lattice to minimize
the Coulomb energy, while the translational energy of nuclei is somewhat small
and can be neglected. Thus, it is reasonable to assume that a unit cell is
periodically repeated in space at very low temperature.
In fact, a periodic structure of the nucleon distribution could be observed
up to $T \sim 3$ MeV in quantum molecular dynamics simulations
for supernova matter~\cite{Wata04,Wata08}.
However, at higher temperature like $T \sim 10$ MeV, the Coulomb lattice
would not survive and the contribution from the translational motion of
nuclei should be properly taken into account.
In recent studies on the equation of state (EOS) for core-collapse supernova
simulations~\cite{Hemp10,Hemp12,Radu10,Gulm15,Furu11,Furu13,Botv13},
the stellar matter at subnuclear densities was described as an ensemble
of nuclei and interacting nucleons in nuclear statistical equilibrium,
where theoretical and experimental nuclear mass tables have been employed
and the translational free energy has been calculated from
Maxwell--Boltzmann statistics.
The distribution of nuclear species can be obtained by minimizing the
total free energy of the system, which is known to be important
for electron captures on nuclei inside supernova core.
However, as shown in Ref.~\cite{Furu11},
the thermodynamic quantities obtained in the nuclear statistical
equilibrium model are not very different from those of the commonly used
single-nucleus approximation, in which only a single representative
nucleus is included.
In the present work, we use the single-nucleus approximation
instead of considering an ensemble of nuclear species and neglect
the translational motion of nuclei for simplicity.

Recently there is an increasing interest in the nuclear symmetry energy
and its density dependence because of their importance for understanding many
phenomena in nuclear physics and astrophysics~\cite{LiBA08,Pal10b,Oyam07,Duco10}.
The symmetry energy $E_{\rm sym}$ at saturation density
is constrained by various experiments to be around $30\pm 4$ MeV,
while the symmetry energy slope $L$ at saturation density
is still quite uncertain and may vary from about $20$ to $115$
MeV~\cite{Chen13}. The influence of the symmetry energy and its slope
on nuclear liquid-gas phase transition was extensively discussed in
bulk calculations~\cite{Sero95,Pal10a,Chen07,Jiang13},
where the two-phase coexistence is governed by the Gibbs conditions
and the surface and Coulomb energies are neglected.
In this study, we aim to investigate the
impact of the symmetry energy on the liquid-gas phase transition
of stellar matter with the inclusion of surface and Coulomb effects.
For the nuclear interaction, we employ the relativistic mean-field (RMF)
theory, which has achieved great success in describing various phenomena
in nuclear physics over the past decades~\cite{Sero97,Bogu77,Ring90,Meng06}.
The RMF theory has recently been reinterpreted by the relativistic
Kohn-Sham density functional theory, which was widely employed
in the treatment of the quantum many-body problem in atomic,
molecular, and condensed matter physics.
In the RMF approach, nucleons interact via the exchange of
isoscalar scalar and vector mesons ($\sigma$ and $\omega$)
and an isovector vector meson ($\rho$), while the parameters
are generally fitted to nuclear matter saturation properties
or ground-state properties of finite nuclei.
We consider two different RMF parametrizations, TM1~\cite{TM1} and
IUFSU~\cite{IUFSU}, which are known to be successful in describing
the ground-state properties of finite nuclei and maximum neutron-star
mass $\sim 2 M_\odot$. The TM1 model was successfully
applied to construct the equation of state for supernova simulations
and neutron stars~\cite{Shen02,Shen11,Zhang14}. The IUFSU model was proposed
to overcome a smaller neutron-star mass predicted by the FSU
model~\cite{FSU}, and meanwhile it could keep an excellent description
of ground-state properties and collective excitations of closed-shell
nuclei~\cite{IUFSU}. These two models include nonlinear terms for
both $\sigma$ and $\omega$ mesons, while an additional $\omega$-$\rho$
coupling term is added in the IUFSU model. It was found that
the $\omega$-$\rho$ coupling term plays an important role in modifying
the density dependence of the symmetry energy and affecting the neutron
star properties~\cite{Horo01,Horo03,Mene11,Prov13}.
To examine the influence of the symmetry energy slope $L$ on
the liquid-gas phase transition of stellar matter,
we adopt two sets of generated models based on the TM1 and IUFSU
parametrizations, which was described in our previous work~\cite{Bao14b}.
These models have been generated by simultaneously adjusting $g_{\rho}$
and ${\Lambda}_{\rm{v}}$ so as to achieve a given $L$ at saturation
density $n_0$ while keeping $E_{\rm{sym}}$ fixed at a density of
0.11 fm$^{-3}$. We note that all models in each set have the same
isoscalar saturation properties and fixed symmetry energy at a density
of 0.11 fm$^{-3}$ but have different symmetry energy slope $L$.
Therefore, these models are ideal for studying the influence of $L$
on the phase transition of stellar matter at subnuclear densities.

This article is organized as follows. In Sec.~\ref{sec:2},
we briefly describe the RMF model and the treatment of a liquid-gas
mixed phase with the inclusion of surface and Coulomb contributions.
In Sec.~\ref{sec:3}, we present the numerical results
and discuss the finite-size effects and the influence of the symmetry
energy on the liquid-gas phase transition of stellar matter.
Section~\ref{sec:4} is devoted to the conclusions.

\section{Formalism}
\label{sec:2}

In this section, we first give a brief description of the RMF theory
adopted for the nuclear interaction. Then, we derive the equilibrium
conditions for two-phase coexistence by using the CLD model, in which
the surface and Coulomb energies are included and calculated self-consistently.
In the RMF approach, nucleons interact via the exchange of various mesons.
The mesons considered are isoscalar scalar and vector mesons ($\sigma$ and
$\omega$) and the isovector vector meson ($\rho$).
The nucleonic Lagrangian density reads
\begin{eqnarray}
\label{eq:LRMF}
\mathcal{L}_{\rm{RMF}} & = & \sum_{i=p,n}\bar{\psi}_i
\left[ i\gamma_{\mu}\partial^{\mu}-\left(M+g_{\sigma}\sigma\right)
-\gamma_{\mu} \left(g_{\omega}\omega^{\mu} +\frac{g_{\rho}}{2}
\tau_a\rho^{a\mu}\right)\right]\psi_i  \notag \\
&& +\frac{1}{2}\partial_{\mu}\sigma\partial^{\mu}\sigma -\frac{1}{2}%
m^2_{\sigma}\sigma^2-\frac{1}{3}g_{2}\sigma^{3} -\frac{1}{4}g_{3}\sigma^{4}
\notag \\
&& -\frac{1}{4}W_{\mu\nu}W^{\mu\nu} +\frac{1}{2}m^2_{\omega}\omega_{\mu}%
\omega^{\mu} +\frac{1}{4}c_{3}\left(\omega_{\mu}\omega^{\mu}\right)^2  \notag
\\
&& -\frac{1}{4}R^a_{\mu\nu}R^{a\mu\nu} +\frac{1}{2}m^2_{\rho}\rho^a_{\mu}%
\rho^{a\mu} +\Lambda_{\rm{v}} \left(g_{\omega}^2
\omega_{\mu}\omega^{\mu}\right)
\left(g_{\rho}^2\rho^a_{\mu}\rho^{a\mu}\right),
\end{eqnarray}
where $W^{\mu\nu}$ and $R^{a\mu\nu}$ are the antisymmetric field
tensors for $\omega^{\mu}$ and $\rho^{a\mu}$, respectively.
In the RMF approach, the meson fields are treated as classical
fields and the field operators are replaced by their expectation
values. For a static system, the nonvanishing expectation values are
$\sigma =\left\langle \sigma \right\rangle$, $\omega =\left\langle
\omega^{0}\right\rangle$, and $\rho =\left\langle \rho^{30} \right\rangle$.
From the Lagrangian density (\ref{eq:LRMF}), we derive the equations
of motion for these mean fields, which can be solved self-consistently.

For uniform nuclear matter at finite temperature, the energy density
is given by
\begin{eqnarray}
\label{eq:ERMF}
\varepsilon &=& \displaystyle{\sum_{i=p,n} \frac{1}{\pi^2}
  \int_0^{\infty} dk\,k^2\,
  \sqrt{k^2+{M^{\ast}}^2}\left( f_{i+}^{k}+f_{i-}^{k}\right) } \nonumber\\
 & &
  +\frac{1}{2}m_{\sigma}^2\sigma^2+\frac{1}{3}g_{2}\sigma^{3}
  +\frac{1}{4}g_{3}\sigma^{4} \nonumber\\
 & &
  +\frac{1}{2}m_{\omega}^2\omega^2+\frac{3}{4}c_{3}\omega^{4}
  +\frac{1}{2}m_{\rho}^2\rho^2
  +3 \Lambda_{\rm{v}}\left(g^2_{\omega}\omega^2\right)
     \left(g^2_{\rho}\rho^2\right),
\end{eqnarray}
the entropy density is written as
\begin{eqnarray}
\label{eq:SRMF}
s &=& -\displaystyle{\sum_{i=p,n}\frac{1}{\pi^{2}}
  \int_{0}^{\infty}dk\,k^{2}}
  \left[ f_{i+}^{k}\ln f_{i+}^{k}+\left( 1-f_{i+}^{k}\right)
  \ln \left(1-f_{i+}^{k}\right) \right.  \nonumber \\
& & \left. +f_{i-}^{k}\ln f_{i-}^{k}
  +\left( 1-f_{i-}^{k}\right) \ln \left( 1-f_{i-}^{k}\right) \right],
\end{eqnarray}
and the pressure is given by
\begin{eqnarray}
\label{eq:PRMF}
 P &=& \displaystyle{\sum_{i=p,n} \frac{1}{3\pi^2}
   \int_0^{\infty} dk\,k^2\,
   \frac{k^2}{\sqrt{k^2+{M^{\ast}}^2}}
   \left( f_{i+}^{k}+f_{i-}^{k}\right) } \nonumber\\
 & &
  -\frac{1}{2}m_{\sigma}^2\sigma^2-\frac{1}{3}g_{2}\sigma^{3}
  -\frac{1}{4}g_{3}\sigma^{4} \nonumber\\
 & &
  +\frac{1}{2}m_{\omega}^2\omega^2+\frac{1}{4}c_{3}\omega^{4}
  +\frac{1}{2}m_{\rho}^2\rho^2
  +\Lambda_{\rm{v}}\left(g^2_{\omega}\omega^2\right)
   \left(g^2_{\rho}\rho^2\right).
\end{eqnarray}
Here $M^{\ast}=M+g_{\sigma}\sigma$ is the effective nucleon mass.
$f_{i+}^{k}$ and $f_{i-}^{k}$ ($i=p,n$) are the occupation
probabilities of nucleon and antinucleon at momentum $k$,
which are given by the Fermi-Dirac distribution,
\begin{eqnarray}
\label{eq:firmf}
f_{i\pm}^{k}=\left\{1+\exp \left[ \left( \sqrt{k^{2}+{M^{\ast}}^2}
 +g_{\omega}\omega+\frac{g_{\rho}}{2}\tau_i^{3}\rho \mp \mu_{i}\right)/T\right]
 \right\}^{-1}.
\end{eqnarray}
The number density of protons ($i=p$) or neutrons ($i=n$) is calculated by
\begin{equation}
\label{eq:nirmf}
 n_{i}=\frac{1}{\pi^2}
       \int_0^{\infty} dk\,k^2\,\left(f_{i+}^{k}-f_{i-}^{k}\right).
\end{equation}

The Wigner--Seitz approximation is adopted to describe inhomogeneous
stellar matter at subnuclear densities. The Wigner--Seitz cell, which consists
of protons, neutrons, and electrons, is assumed to be spherical and charge
neutral. We consider only droplet and bubble configurations,
while other pasta phases in between are neglected.
In the present work, we employ the single-nucleus approximation
and neglect the translational motion of nuclei for simplicity.
The CLD model is used to describe the nucleus in the Wigner--Seitz cell.
The matter inside the cell is assumed to separate into a dense liquid ($L$)
phase and a dilute gas ($G$) phase with a sharp interface.
We assume a uniform distribution of electrons throughout the Wigner-Seitz
cell because the electron screening effect is known to be negligible at
subnuclear densities~\cite{Maru05}. At given temperature $T$, average
baryon density $n_b$, and proton fraction $Y_p$, the equilibrium state
should be determined by minimizing the total free energy density of
the system~\cite{Latt91,Shen11,Bao14b}.
We consider both droplet and bubble configurations, where the free energy
density of the cell is written as
\begin{equation}
\label{eq:fws}
f=u f^L \left(n^L_{p},n^L_{n}\right)
    +\left(1-u\right)f^G\left(n^G_{p},n^G_{n}\right)
    +f_e\left(n_e\right)
    +f_{\rm{surf}}\left(u,r_d,\tau\right)
    +f_{\rm{Coul}}\left(u,r_d,n^L_{p},n^G_{p}\right).
\end{equation}
Here, $u$ is the volume fraction of the liquid phase, $r_d$ is
the radius of the droplet or bubble, and $\tau$ is the surface tension.
The proton and neutron densities
in the liquid (gas) phase are denoted by $n^L_{p}$ ($n^G_{p}$) and
$n^L_{n}$ ($n^G_{n}$), respectively. The free energy density
of uniform nuclear matter in phase $i$ ($i=L,G$),
$f^i=\varepsilon^i-T s^i$,
can be obtained in the RMF approach with the energy and entropy densities
given by Eqs.~(\ref{eq:ERMF}) and (\ref{eq:SRMF}).
The surface and Coulomb terms for a spherical cell
are given by
\begin{eqnarray}
{f}_{\rm{surf}}
&=& \frac{3 \tau u_{\rm{in}}}{r_d},
\label{eq:esurf} \\
{f}_{\rm{Coul}}
&=& \frac{e^2}{5}
    \left(n^L_p-n^G_p\right)^{2}r_d^{2} u_{\rm{in}} D\left( u_{\rm{in}}\right),
\label{eq:ecoul}
\end{eqnarray}%
with%
\begin{eqnarray}
\label{eq:Du}
D\left( u_{\rm{in}}\right) =1-\frac{3}{2}u_{\rm{in}}^{1/3}+\frac{1}{2}u_{\rm{in}}.
\end{eqnarray}%
Here, $u_{\rm{in}}$ denotes the volume fraction of the inner part,
so we have $u_{\rm{in}}=u$ for droplets and $u_{\rm{in}}=1-u$ for bubbles.
$e=\sqrt{4\pi/137}$ is the electromagnetic coupling constant.
The surface tension $\tau$ is calculated by using a
Thomas-Fermi approach for a one-dimensional nuclear system with the same
RMF parametrization~\cite{Bao14a,Mene10,Agra14}.
As discussed in Refs.~\cite{Agra14,Cent98}, there could be two definitions
of the nuclear interface energy. We calculate the surface tension $\tau$
from the thermodynamic potential per unit area~\cite{Agra14},
\begin{equation}
\tau=\int_{-\infty}^{\infty}dz\left\{f(z)-f^{G}
-\mu_p\left[n_p(z)-n^{G}_{p}\right]
-\mu_n\left[n_n(z)-n^{G}_{n}\right]\right\},
\label{eq:tau}
\end{equation}
where both the surface energy and the surface entropy are included
in the first term. The equilibrium density profiles, $n_p(z)$ and $n_n(z)$,
can be obtained in the Thomas-Fermi approach at finite temperature.

It is clear that the free energy density $f$ given by Eq.~(\ref{eq:fws})
is a function of seven variables:
$n^L_{p}$, $n^L_{n}$, $n^G_{p}$, $n^G_{n}$, $n_e$, $u$, and $r_d$.
For a charge neutral system with fixed temperature $T$, average baryon
density $n_b$, and proton fraction $Y_p$, these seven variables
are not independent and they should satisfy the following constraints:
\begin{eqnarray}
u n^L_{p} + \left(1-u\right) n^G_{p} &=& n_b Y_p,  \label{eq:cnp} \\
u n^L_{n} + \left(1-u\right) n^G_{n} &=& n_b \left(1-Y_p\right), \label{eq:cnn} \\
n_e &=& n_b Y_p.  \label{eq:cne}
\end{eqnarray}
To derive the phase equilibrium conditions by minimizing the
free energy density of the cell, we introduce the Lagrange multipliers,
$\mu_p$, $\mu_n$, and $\mu_e$, for these constraints, and perform the
minimization for the function,
\begin{eqnarray}
w=f- \mu_p \left[ u n^L_{p} + \left(1-u\right) n^G_{p}\right]
        - \mu_n \left[ u n^L_{n} + \left(1-u\right) n^G_{n}\right]
        - \mu_e n_e.
\end{eqnarray}
Minimizing $w$ with respect to the variables yields the following results:
\begin{eqnarray}
 0 &=& \frac{\partial w}{\partial n^L_{n}}
   = u \left[ \frac{\partial f^L }{\partial n^L_{n}}-\mu_{n}\right]
   = u \left[ \mu_{n}^{L}-\mu_{n} \right],
\label{eq:min1} \\
 0 &=& \frac{\partial w}{\partial n^G_{n}}
   = (1-u) \left[ \frac{\partial f^G }
     {\partial n^G_{n}}-\mu_{n}\right]
   = (1-u) \left[ \mu_{n}^{G}-\mu_{n} \right],
\label{eq:min2} \\
 0 &=& \frac{\partial w}{\partial n^L_{p}}
   = u \left[ \frac{\partial f^L }
     {\partial n^L_{p}}-\mu_{p}\right]
     +\frac{2 f_{\rm{Coul}} }{n^L_{p}-n^G_{p}}
   = u \left[\mu_{p}^{L}-\mu_{p}\right]
     +\frac{2 f_{\rm{Coul}} }{n^L_{p}-n^G_{p}},
\label{eq:min3} \\
 0 &=& \frac{\partial w}{\partial n^G_{p}}
   = (1-u) \left[ \frac{\partial f^G }
     {\partial n^G_{p}}-\mu_{p}\right]
     -\frac{2 f_{\rm{Coul}} }{n^L_{p}-n^G_{p}}
   = (1-u) \left[ \mu_{p}^{G}-\mu_{p}\right]
     -\frac{2 {f}_{\rm{Coul}} }{n^L_{p}-n^G_{p}},
\label{eq:min4} \\
 0 &=& \frac{\partial w}{\partial u}
   = \left[ f^L-\mu_{p}n_{p}^{L}-\mu_{n}n_{n}^{L}\right]
    -\left[ f^G-\mu_{p}n_{p}^{G}-\mu_{n}n_{n}^{G}\right]
   \pm \left[ \frac{f_{\rm{surf}} }{u_{\rm{in}}}
             +\frac{f_{\rm{Coul}} }{u_{\rm{in}}}
              \left( 1+u_{\rm{in}}\frac{D^{^{\prime }}}{D}\right)\right] ,
\label{eq:min5} \\
 0 &=& \frac{\partial w}{\partial r_d}
   = -\frac{  f_{\rm{surf}} }{r_d}
     +\frac{2 f_{\rm{Coul}} }{r_d}
   = \frac{1}{r_d}\left[2 f_{\rm{Coul}}-f_{\rm{surf}} \right].
\label{eq:min6}
\end{eqnarray}
Note that electrons play no role in the minimization procedure
because the electron density was fixed according to Eq.~(\ref{eq:cne}).
For simplicity, we have neglected contributions from the derivatives of
the surface tension in deriving the above equations~\cite{Bao14b}.
From Eqs.~(\ref{eq:min1})--(\ref{eq:min6}) we can obtain the equilibrium
conditions between liquid and gas phases in droplet and bubble
configurations,
\begin{eqnarray}
\mu_{n}^{G} &=& \mu_{n}^{L},
\label{eq:cmun} \\
\mu_{p}^{G} &=&
\mu_{p}^{L}+\frac{2f_{\rm{Coul}} }{u(1-u)\left(n^L_{p}-n^G_{p}\right)},
\label{eq:cmup} \\
P^{G} &=&
P^{L}+\frac{2f_{\rm{Coul}} }{n^L_{p}-n^G_{p}}
  \left( \frac{n_p^L}{u}+\frac{n_p^G}{1-u} \right)
  \mp \frac{f_{\rm{Coul}} }{u_{\rm{in}}}
  \left(3+u_{\rm{in}}\frac{D^{^{\prime }}}{D}\right).
\label{eq:cp}
\end{eqnarray}%
In Eqs.~(\ref{eq:min5}) and (\ref{eq:cp}), the sign of the last term is
\textquotedblleft $-$\textquotedblright\ for droplets
and \textquotedblleft $+$\textquotedblright\ for bubbles.
The pressure of uniform nuclear matter in phase $i$ ($i=L,G$)
is given by $P^i=\mu_{p}^{i}n_{p}^{i}+\mu_{n}^{i}n_{n}^{i}-f^i$.
We have checked that Eqs.~(\ref{eq:cmun})--(\ref{eq:cp}) are consistent
with the equilibrium equations given in Refs.~\cite{Bao14b,Wata00,Chamel08}.
One can see that Eqs.~(\ref{eq:cmun})--(\ref{eq:cp}) reduce back to
Gibbs equilibrium conditions when the surface and Coulomb terms
are neglected. It is clear that equilibrium conditions for two-phase
coexistence are significantly altered because of the inclusion of
surface and Coulomb contributions in the minimization procedure.

By solving the above equilibrium equations
at given temperature $T$, average baryon density $n_b$,
and proton fraction $Y_p$,
we can obtain the properties of the two coexisting phases,
and then calculate thermodynamic quantities of the mixed phase.
Based on the equilibrium condition
$f_{\rm{surf}}=2 f_{\rm{Coul}}$ obtained from
Eq.~(\ref{eq:min6}), the radius of the droplet or bubble is given by
\begin{eqnarray}
\label{eq:RD}
r_d &=& \left[\frac{15\tau}
        {2 e^2 \left(n^L_p-n^G_p\right)^{2} D} \right]^{1/3}.
\end{eqnarray}
Then, the radius of the Wigner--Seitz cell is obtained from
$r_{\rm{ws}} = u_{\rm{in}}^{-1/3} r_d$.
In practice, we solve the coupled Eqs.~(\ref{eq:cmun})--(\ref{eq:cp})
together with the meson-field equations in the two coexisting phases
obtained from the RMF model. At finite temperature, the mixed phase
exists only over a finite range of density. Therefore, no solution
can be found at very low and very high densities.
It is very interesting to investigate the finite-size effect on
the boundary of the liquid-gas coexistence region.

\section{Results and discussion}
\label{sec:3}

In this section, we investigate the finite-size effect and the influence
of the symmetry energy on the liquid-gas phase transition of stellar matter.
For the nuclear interaction, we employ two successful RMF models,
TM1~\cite{TM1} and IUFSU~\cite{IUFSU}. The parameter sets and
saturation properties of these two models are given in Tables~\ref{tab:1}
and~\ref{tab:2}, respectively. One can see that the TM1 model predicts
very large symmetry energy $E_{\text{sym}}$ and its slope $L$ at
saturation density, while those of the IUFSU model are relatively small.
It is well known that the symmetry energy slope $L$ plays an important
role in determining the neutron-skin thickness of finite nuclei and
various properties of neutron stars~\cite{Horo01,Mene11,Bao15}.
To clarify the influence of the symmetry energy slope $L$
on the liquid-gas phase transition of stellar matter,
we employ two sets of generated
models based on the TM1 and IUFSU parametrizations,
which were obtained in Ref.~\cite{Bao14b} by simultaneously adjusting
$g_{\rho}$ and ${\Lambda}_{\rm{v}}$ to achieve a given $L$ at
saturation density and keep $E_{\rm{sym}}$ fixed at a density of
$0.11\, \rm{fm}^{-3}$. The resulting parameters, $g_{\rho}$ and
${\Lambda}_{\rm{v}}$, have been presented in
Tables $\textrm{\uppercase\expandafter{\romannumeral2}}$
and $\textrm{\uppercase\expandafter{\romannumeral3}}$ of
Ref.~\cite{Bao14b}.
It is noticeable that all models in each set have
the same isoscalar saturation properties and fixed symmetry energy
$E_{\rm{sym}}$ at a density of $0.11\, \rm{fm}^{-3}$ but have different
symmetry energy slope $L$.

\subsection{Finite-size effects}
\label{sec:3-1}

\begin{figure}[htb]
\includegraphics[bb=19 49 565 787, width=7 cm,clip]{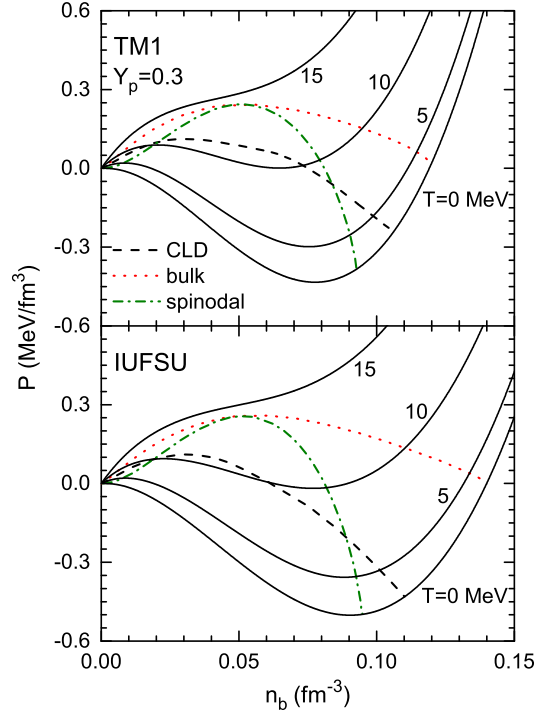}
\caption{(Color online) Pressure of uniform matter $P$
as a function of baryon density $n_b$ at fixed proton fraction $Y_p=0.3$
for various temperature $T$ obtained in the TM1 (upper panel)
and IUFSU (lower panel) models. The black dashed lines represent
the coexistence region obtained with finite-size effects
in the CLD model. The red dotted lines represent the coexistence
region obtained from a bulk calculation.
The green dashed-dotted lines indicate the spinodal region
determined by the curvature matrix of the free energy. }
\label{fig:1Pnb}
\end{figure}

\begin{figure}[htb]
\includegraphics[bb=17 18 548 759, width=7 cm,clip]{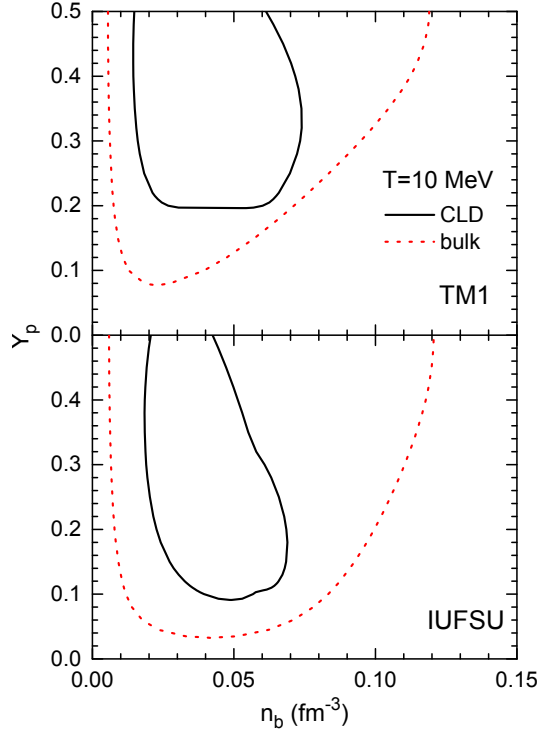}
\caption{(Color online) Phase diagram in the $n_b$-$Y_p$ plane
at $T=10$ MeV obtained in the TM1 (upper panel) and IUFSU (lower panel) models.
The black solid lines indicate the boundaries of
the coexistence region obtained with finite-size effects in the CLD model,
while the red dotted lines correspond to the results obtained from
a bulk calculation.}
\label{fig:2Ypnb}
\end{figure}

\begin{figure}[htb]
\includegraphics[bb=23 18 541 758, width=7 cm,clip]{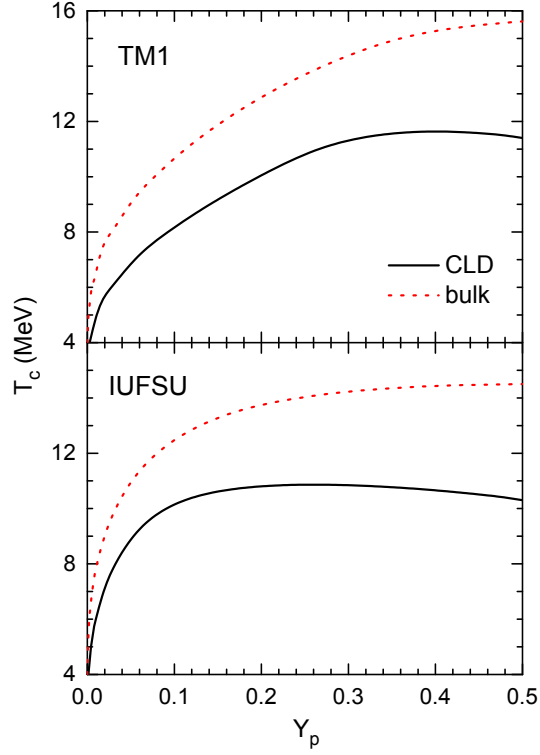}
\caption{(Color online) Critical temperature $T_c$ as a function of
proton fraction $Y_p$ obtained in the TM1 (upper panel) and
IUFSU (lower panel) models.
The results of the CLD model with finite-size effects are
indicated by the black solid lines, while those from a bulk calculation
are indicated by the red dotted lines.}
\label{fig:3TcYp}
\end{figure}

We first investigate finite-size effects on the boundary of the liquid-gas
coexistence region of stellar matter. In this study, we consider both
droplet and bubble configurations. In the mixed phase, droplets appear
at low density while bubbles are formed at high density.
In Fig.~\ref{fig:1Pnb}, we show the pressure of uniform matter as
a function of the baryon density $n_b$ with fixed proton fraction $Y_p=0.3$
at various temperature $T$.
Results of the TM1 and IUFSU models are presented in the
upper and lower panels, respectively. The dashed-dotted line indicates
the boundary of the spinodal region which is determined
by the curvature matrix of the free energy as described in Ref.~\cite{Marg03}.
The phase coexistence region (binodal curve) obtained with finite-size
effects in the CLD model is shown by the dashed line, while the one
obtained in a bulk calculation is shown by the dotted line.
By comparing the dashed and dotted lines, we see that the inclusion of
surface and Coulomb contributions can significantly reduce the phase coexistence
region. The binodal curve with finite-size effects (dashed line) is even
lower than the spinodal curve (dashed-dotted line) at higher temperatures.
This is because after adding surface and Coulomb contributions,
the free energy of the mixed phase becomes higher than that of the single phase
in the spinodal instability region, so the thermodynamically favorable state
is the single phase in this case.
It is well known that properties of nuclear liquid-gas phase
transition are sensitive to the neutron-proton asymmetry.
We show in Fig.~\ref{fig:2Ypnb} the phase diagram in the $n_b$-$Y_p$ plane
at $T=10$ MeV obtained in the TM1 (upper panel) and IUFSU (lower panel) models.
The solid and dotted lines, respectively, indicate the boundaries of the
liquid-gas coexistence region calculated with and without finite-size effects.
It is evident that the coexistence region obtained with surface and Coulomb
contributions is much smaller than that obtained from the bulk calculation.
Furthermore, the isospin symmetry of the nuclear system is broken by Coulomb
interaction, so that the maximum density range of the mixed phase is not
achieved at $Y_p=0.5$ when the contributions from Coulomb and surface terms
are taken into account in the CLD model.
By comparing the two panels of Fig.~\ref{fig:2Ypnb},
one can see that the smallest $Y_p$ for two-phase coexistence obtained in
the TM1 model is somewhat larger than that obtained in the IUFSU model.
This is because the TM1 model has a much larger value of the symmetry
energy slop $L$ than the IUFSU model (see Table~\ref{tab:2}).
The correlation between $L$ and the smallest $Y_p$ will be discussed
in Sec.~\ref{sec:3-2}. For the liquid-gas coexistence phase,
there is a critical temperature $T_c$, above which two-phase equilibrium
can not be achieved and only a single phase may exist.
In Fig.~\ref{fig:3TcYp}, we show the critical temperature $T_c$ as
a function of the proton fraction $Y_p$ obtained in the TM1 (upper panel)
and IUFSU (lower panel) models. The results with and without finite-size
effects are plotted by solid and dotted lines, respectively.
It is seen that the inclusion of surface and Coulomb contributions results in
a significant decrease of $T_c$. Because of the Coulomb interaction,
the largest $T_c$ with finite-size effects is not achieved at $Y_p=0.5$,
which is different from the results of bulk calculations.
One can see that with decreasing $Y_p$, the decrease of $T_c$ in the TM1
model is more pronounced than that in the IUFSU model,
which may be related to the difference of $L$ in these two models.

\begin{center}
\begin{figure}[htb]
\centering
\begin{tabular}{ccc}
\includegraphics[bb=14 164 564 663, width=0.33\linewidth, clip]{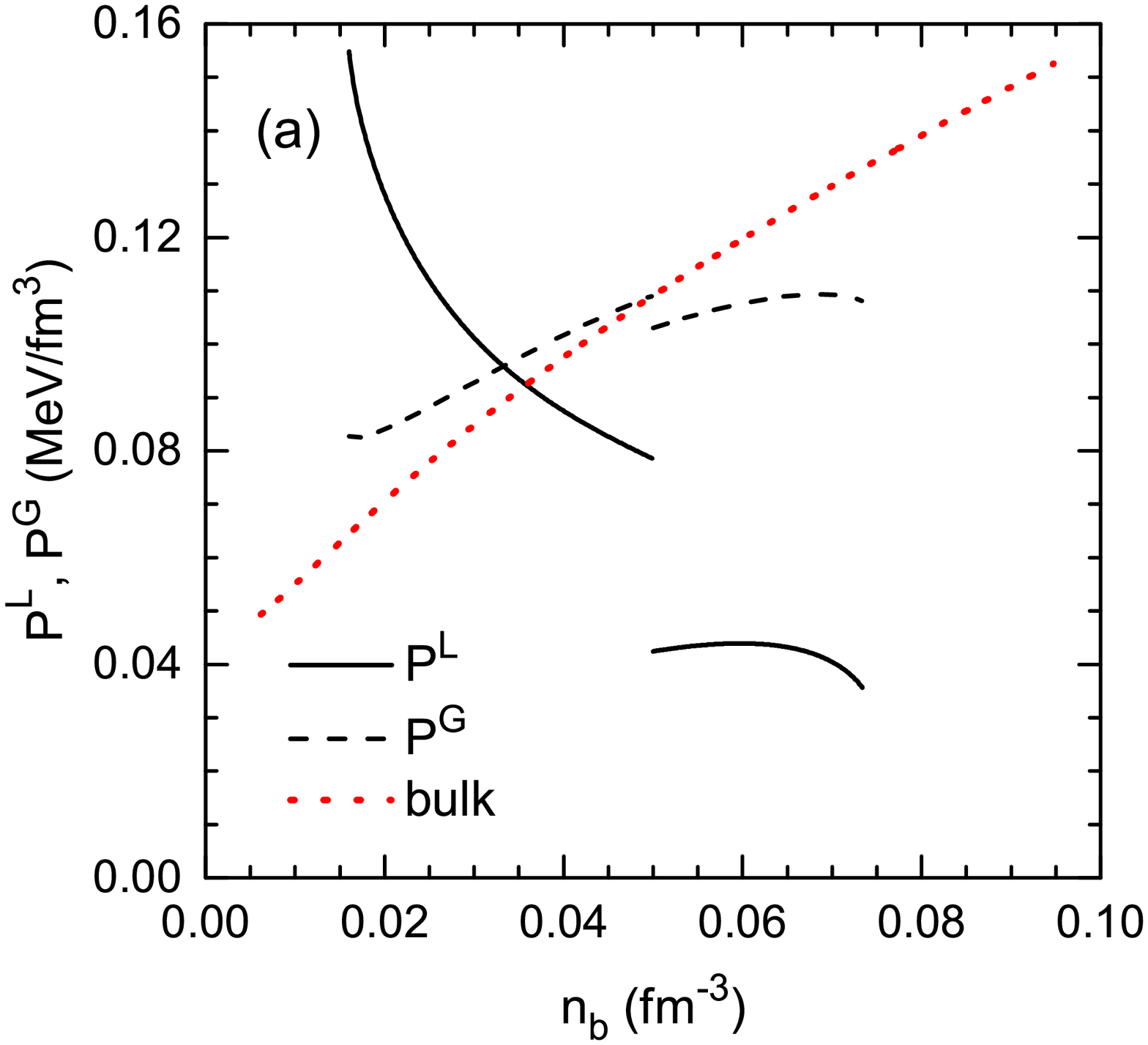}&
\includegraphics[bb=14 164 564 663, width=0.33\linewidth, clip]{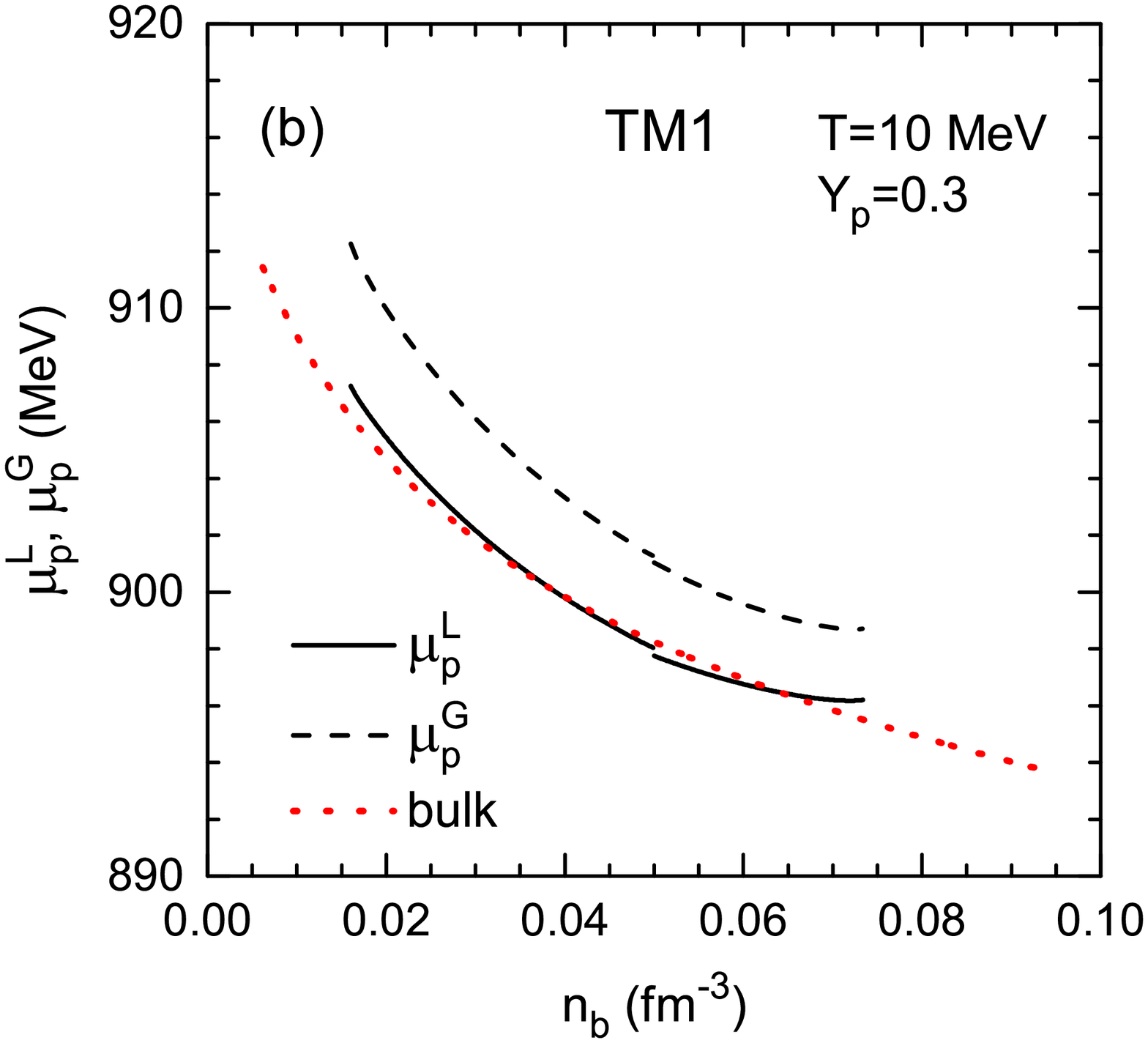}&
\includegraphics[bb=14 164 564 663, width=0.33\linewidth, clip]{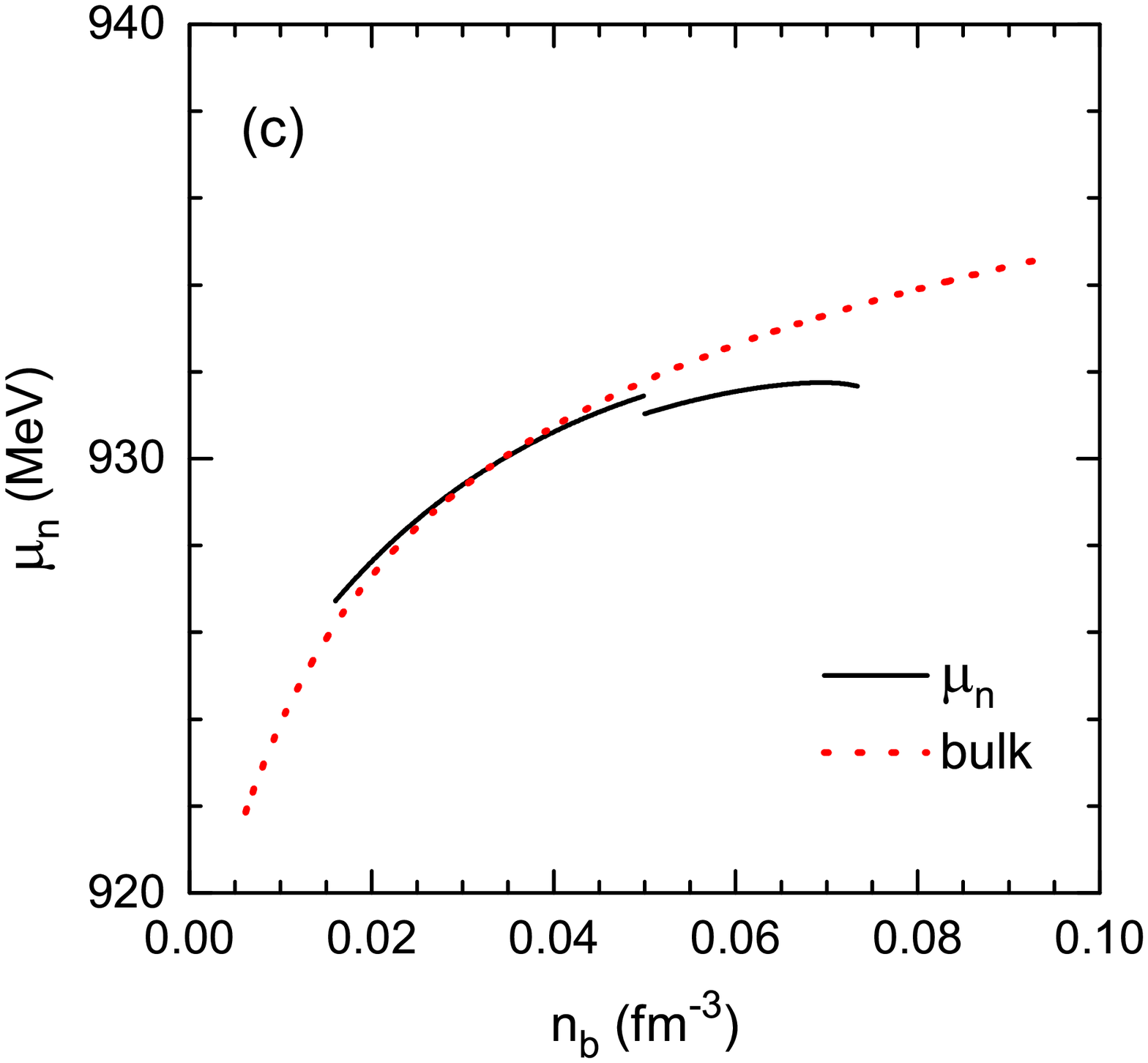}  \\
\includegraphics[bb=14 164 564 663, width=0.33\linewidth, clip]{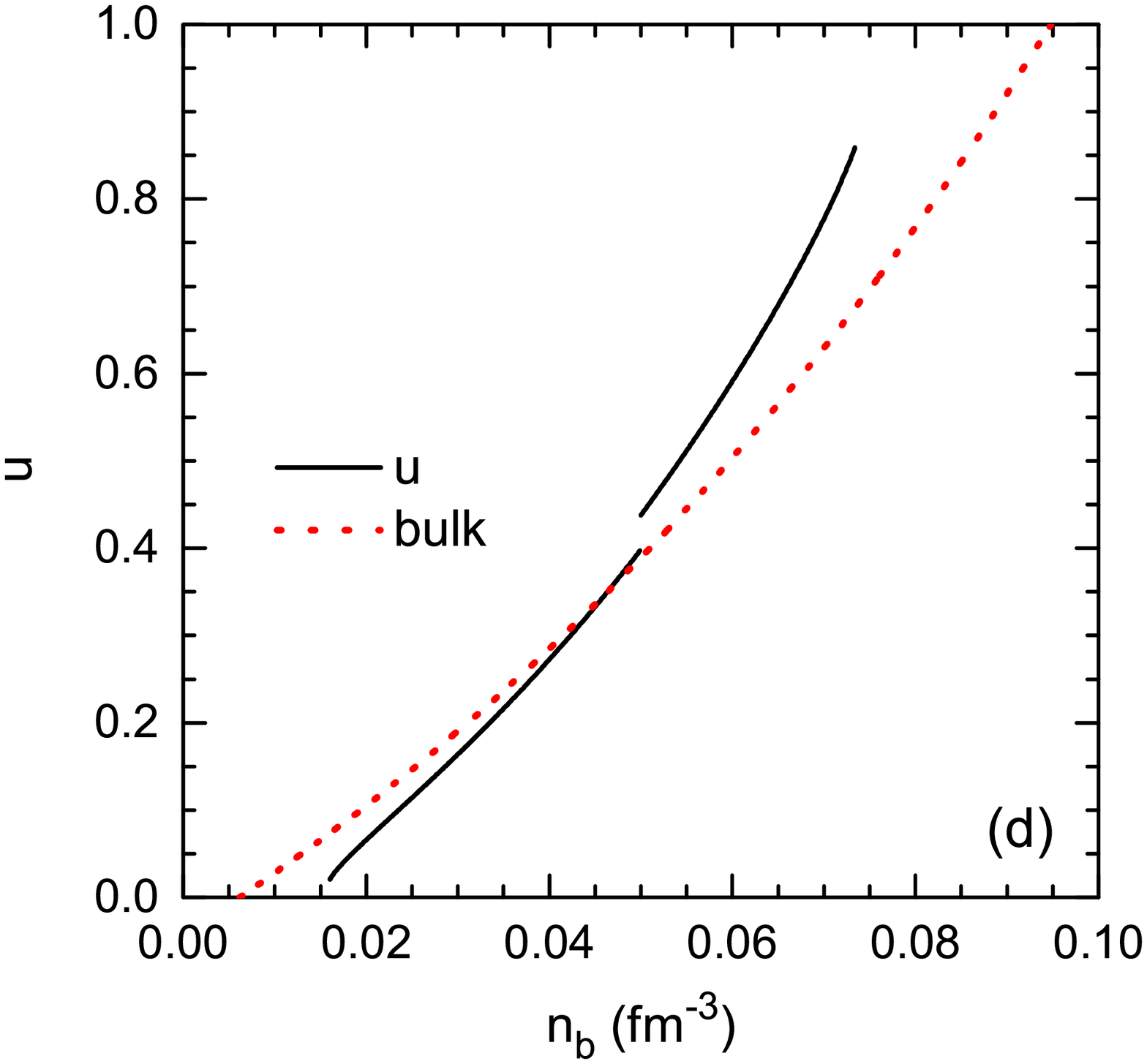}&
\includegraphics[bb=14 164 564 663, width=0.33\linewidth, clip]{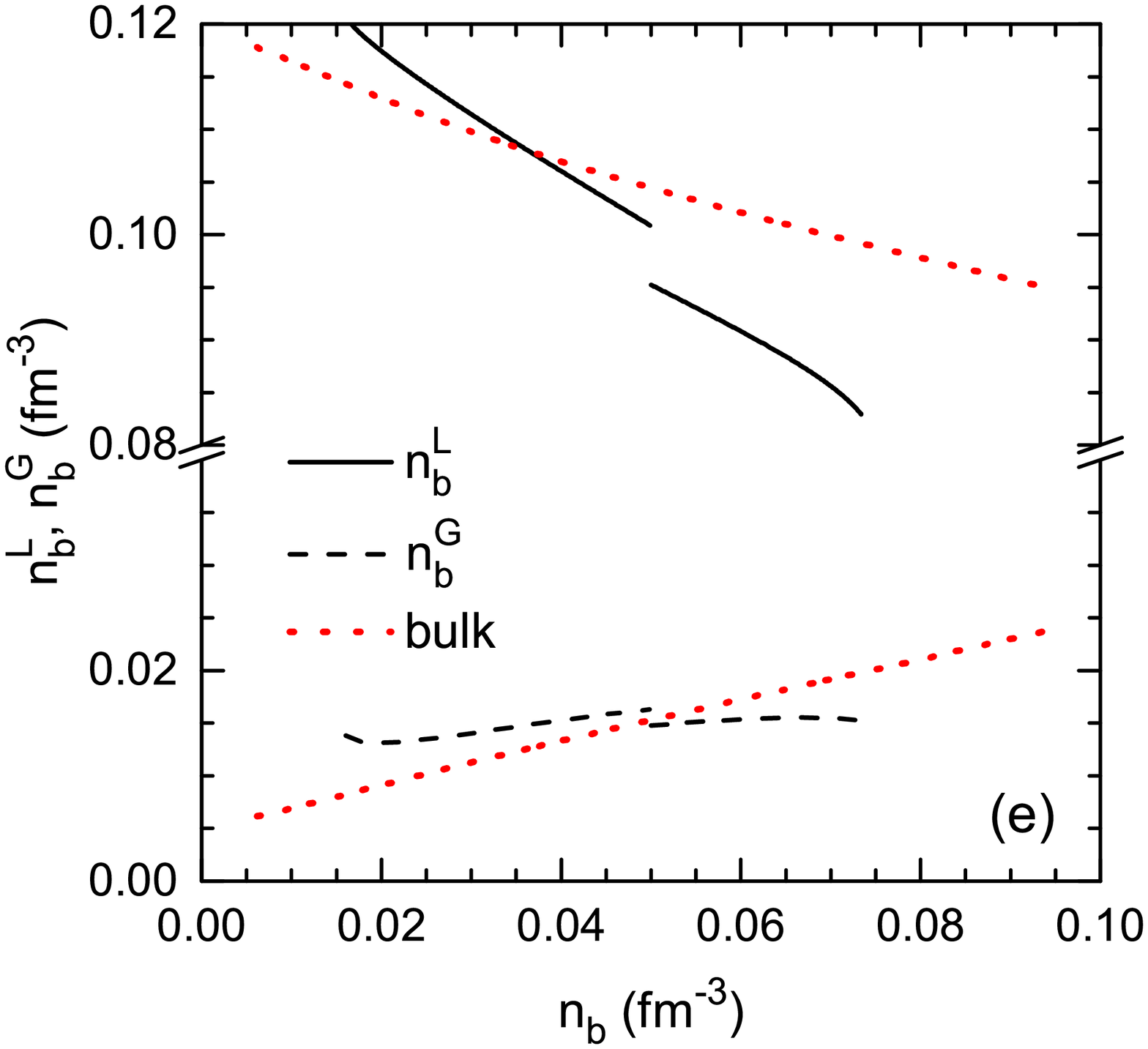}&
\includegraphics[bb=14 164 564 663, width=0.33\linewidth, clip]{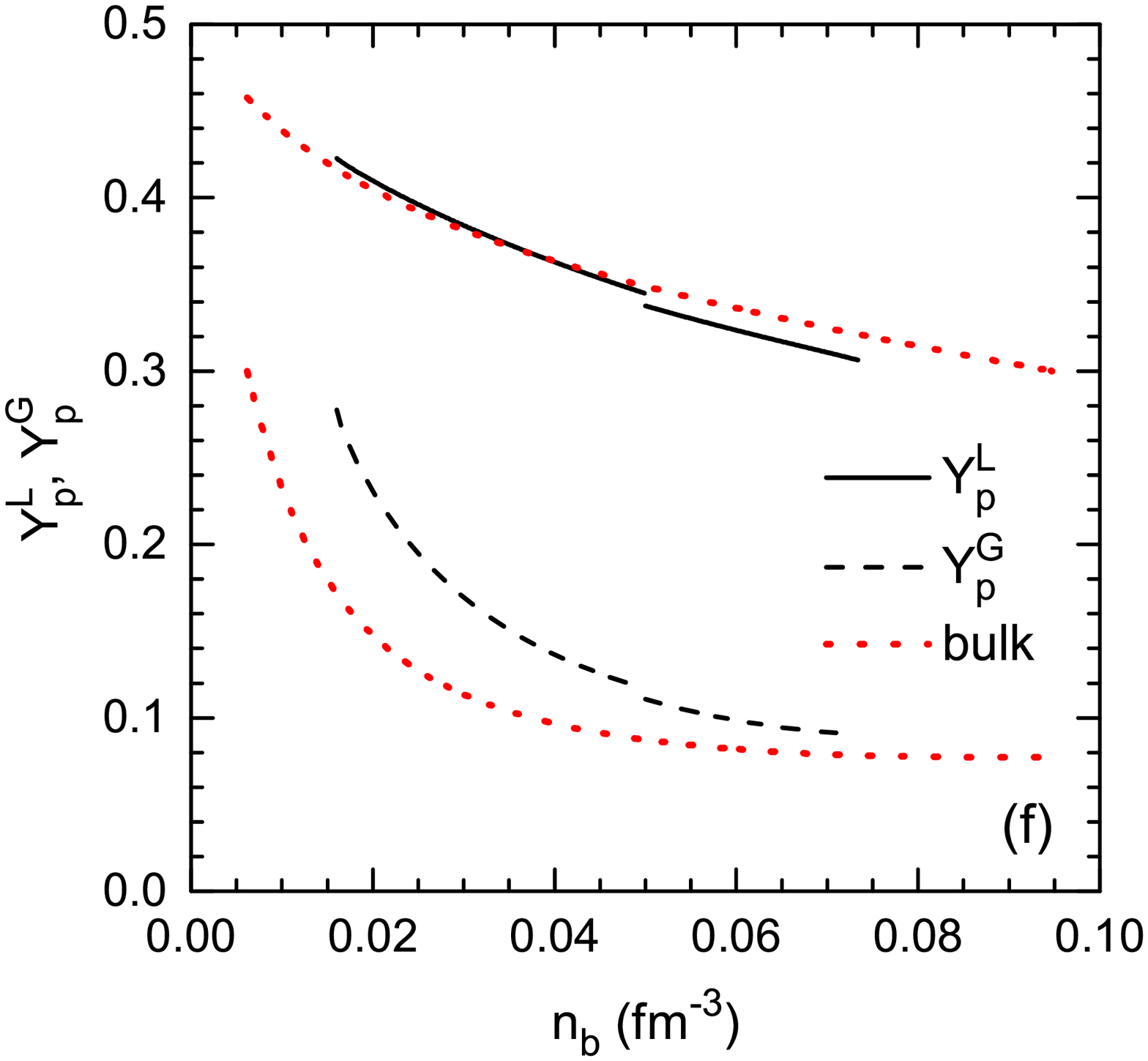} \\
\end{tabular}
\caption{(Color online) Properties of the liquid ($L$) and gas ($G$) mixed
phase at $T=10$ MeV and $Y_p=0.3$ obtained with finite-size effects
using the TM1 model.
Pressures $P^L$ and $P^G$ (a),
proton chemical potentials $\mu^L_p$ and $\mu^G_p$ (b),
neutron chemical potential $\mu_n$ (c),
volume fraction of the liquid phase $u$ (d),
baryon densities $n^L_b$ and $n^G_b$ (e),
and proton fractions $Y^L_p$ and $Y^G_p$ (f)
are plotted as a function of the average baryon density $n_b$.
The corresponding results of a bulk calculation are shown by
the red dotted lines for comparison.}
\label{fig:4Pnb}
\end{figure}
\end{center}

It is interesting to examine the influence of surface and Coulomb
effects on properties of the liquid-gas mixed phase of stellar matter.
In the bulk calculation, the finite-size effects like surface and Coulomb
contributions are neglected and the two coexisting phases are governed
by the Gibbs conditions,
which demand equal pressure and chemical potentials for coexisting phases.
However, when surface and Coulomb contributions are taken into account,
the phase equilibrium conditions obtained by minimizing the total free energy
are given by Eqs.~(\ref{eq:cmun})--(\ref{eq:cp}), which imply the pressure
and the proton chemical potential in the liquid phase are different from
those in the gas phase. Furthermore, other properties of the mixed phase,
such as coexisting densities and proton fractions of the liquid and gas phases,
are also affected by the finite-size effects.
In Fig.~\ref{fig:4Pnb}, we show a comparison between
the results obtained with and without finite-size effects. The calculations
are performed at $T=10$ MeV and $Y_p=0.3$ using the TM1 parametrization.
We plot in Figs.~\ref{fig:4Pnb}(a) and~\ref{fig:4Pnb}(f) the following quantities 
as a function of the average baryon density $n_b$:
(a) pressures $P^L$ and $P^G$;
(b) proton chemical potentials $\mu_p^L$ and $\mu_p^G$;
(c) neutron chemical potential $\mu_n=\mu_n^L=\mu_n^G$;
(d) volume fraction of the liquid phase $u$;
(e) baryon densities $n_b^L$ and $n_b^G$;
(f) proton fractions $Y_p^L$ and $Y_p^G$.
It is noticeable that there are clear discontinuities
at $n_b\sim 0.05\, \rm{fm}^{-3}$ in the results of the CLD model,
which are caused by the transition from droplet to bubble.
Similar discontinuities were also observed between different pasta phases
in the CLD and Thomas--Fermi calculations of Ref.~\cite{Pais15}.
In Fig.~\ref{fig:4Pnb}(a), we can see that
the pressure of the bulk calculation increases monotonically with
increasing $n_b$, but the pressures of the liquid and gas phases,
$P^L$ and $P^G$, obtained with finite-size effects show different behaviors.
In the droplet configuration, $P^L$ decreases and $P^G$ increases as $n_b$
increases, which may be caused by the decrease of the surface tension $\tau$
and by the increase of the liquid volume fraction $u$.
It is seen in Fig.~\ref{fig:4Pnb}(b) that the proton chemical potential
in the gas phase $\mu_p^G$ is larger than the one in the liquid phase
$\mu_p^L$,  while the proton chemical potential obtained in the bulk
calculation is very close to the value of $\mu_p^L$. This is because
the inclusion of Coulomb interaction favors a small difference in the
proton density between the liquid and gas phases, so that $\mu_p^G$
is raised to lower the difference between $n_p^L$ and $n_p^G$.
Because of the same reason, $Y_p^G$ in the CLD model is obviously larger
than that in the bulk calculation, as shown in Fig.~\ref{fig:4Pnb}(f).
On the other hand, differences in the results of droplets
with and without finite-size effects are relatively small
in Figs.~\ref{fig:4Pnb}(c) and~\ref{fig:4Pnb}(d).
The behavior of proton and neutron chemical potentials 
was extensively discussed in Ref.~\cite{Pais15},
where calculations were performed using three different approaches
with the FSU parametrization. Our results shown in Figs.~\ref{fig:4Pnb}(b)
and~\ref{fig:4Pnb}(c) are consistent with their CLD calculations.

\subsection{Symmetry energy effects}
\label{sec:3-2}

\begin{figure}[htb]
\includegraphics[bb=23 18 541 758, width=7 cm,clip]{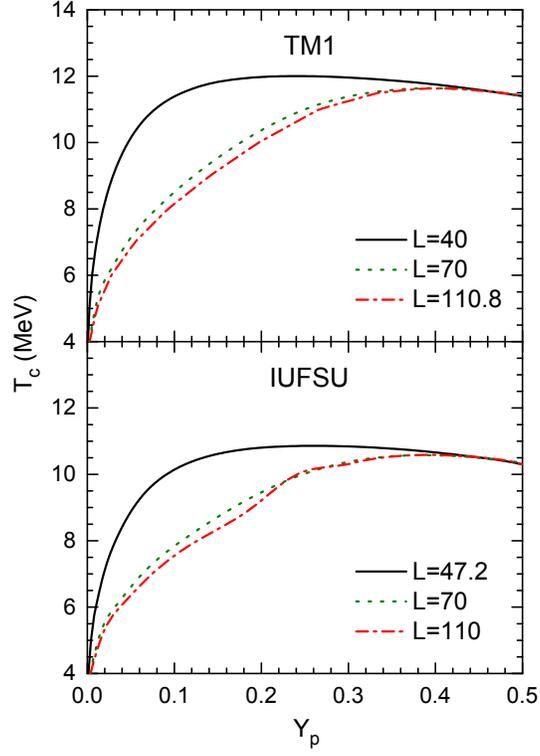}
\caption{(Color online) Critical temperature $T_c$ as a function of
proton fraction $Y_p$ with finite-size effects using generated
models of TM1 (upper panel) and IUFSU (lower panel).}
\label{fig:5LTcYp}
\end{figure}

We explore the effects of the symmetry energy and its slope on properties
of the liquid-gas phase transition of stellar matter.
In previous studies~\cite{Sero95,Pal10a,Chen07,Jiang13}, these effects
have been discussed in the bulk calculations without finite-size effects.
In the present work, we study the symmetry energy effects on the
liquid-gas phase transition with the inclusion of surface and Coulomb
contributions. We employ two sets of generated models based on the TM1 and
IUFSU parametrizations. All models in each set have the same isoscalar
saturation properties and fixed symmetry energy at a density of
$0.11\, \rm{fm}^{-3}$ but have different symmetry energy slope $L$.
In Fig.~\ref{fig:5LTcYp}, we show the critical temperature $T_c$ as
a function of the proton fraction $Y_p$ for the two sets of models
generated from TM1 (upper panel) and IUFSU (lower panel) parametrizations.
One can see that in each panel the models with different $L$ predict
the same value of $T_c$ at $Y_p=0.5$, which is because the differences
of $g_{\rho}$ and ${\Lambda}_{\rm{v}}$ between the models have no effect
on properties of symmetric nuclear matter. However, at small $Y_p$,
there are considerable differences in $T_c$ between the models with
different $L$. The model with a large $L$ predicts a small $T_c$.
At $Y_p=0.1$, the original TM1 model with $L=110.8$ MeV predicts
$T_c\sim 8$ MeV, whereas the generated model with $L=40$ MeV gives
$T_c\sim 11.4$ MeV. At $Y_p=0.3$, the difference in $T_c$ becomes much
less, where $T_c$ is in the range of 11.2--12 MeV (see the upper panel
of Fig.~\ref{fig:5LTcYp}). These results obtained in the CLD model
with the TM1 parametrization are very close to the values of the
Thomas-Fermi calculation shown in Fig.~1 of Ref.~\cite{Shen11}.
The authors of Ref.~\cite{Pais15} have used the FSU parametrization
and compared the results obtained from the CLD model with those
from the Thomas--Fermi approach. They found that different approaches
give very similar results for the crust-core transition densities,
but the CLD model can not predict the existence of the pasta phase at
$T=10$ MeV and $Y_p=0.3$, while the Thomas--Fermi approach predicts
that bubbles exist until $n_b=0.068$ fm$^{-3}$
(see Table~\ref{tab:2} of Ref.~\cite{Pais15}).
This may be from the parametrized surface tension used in their
CLD calculation being too high relative to the value of the Thomas--Fermi
calculation. In the present work, we use the surface tension calculated
from the Thomas-Fermi approach without any additional parametrization.
Within the original TM1 model, we obtained the transition density
to uniform matter is about $0.073$ fm$^{-3}$ at $T=10$ MeV and
$Y_p=0.3$ in the CLD approach, which is close to the corresponding
value of $0.069$ fm$^{-3}$ obtained from the Thomas--Fermi calculation
of Ref.~\cite{Zhang14}.

\begin{figure}[htb]
\includegraphics[bb=27 36 548 783, width=7 cm,clip]{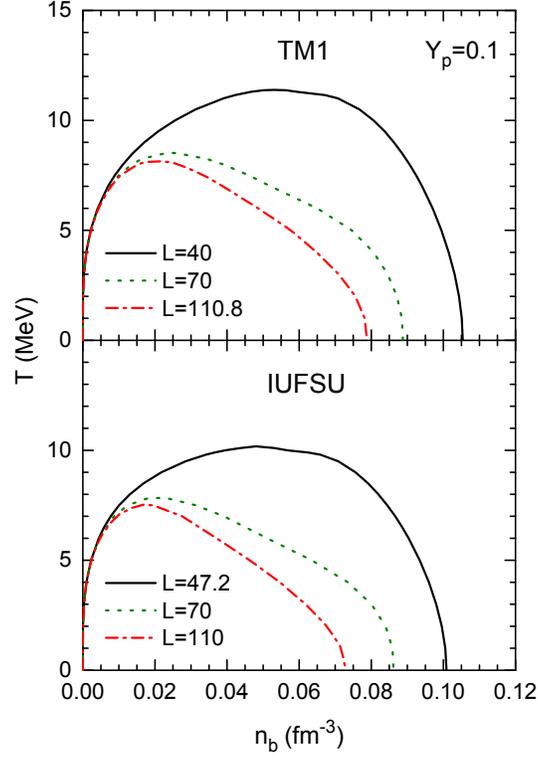}
\caption{(Color online) Phase diagram in the $n_b$-$T$ plane at $Y_p=0.1$
obtained using generated models of TM1 (upper panel) and
IUFSU (lower panel).}
\label{fig:6LTnb}
\end{figure}

\begin{figure}[htb]
\includegraphics[bb=26 37 558 783, width=7 cm,clip]{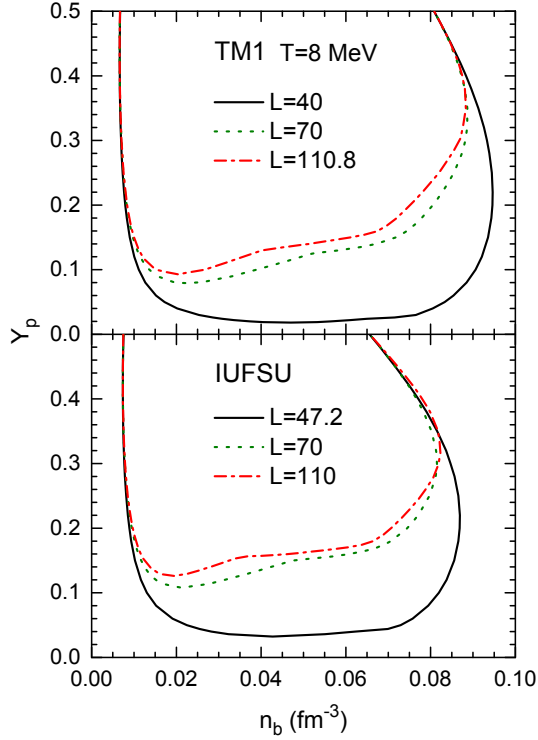}
\caption{(Color online) Phase diagram in the $n_b$-$Y_p$ plane at $T=8$ MeV
obtained using generated models of TM1 (upper panel) and
IUFSU (lower panel).}
\label{fig:7LYpnb}
\end{figure}

\begin{figure}[htb]
\includegraphics[bb=21 165 560 665, width=7 cm,clip]{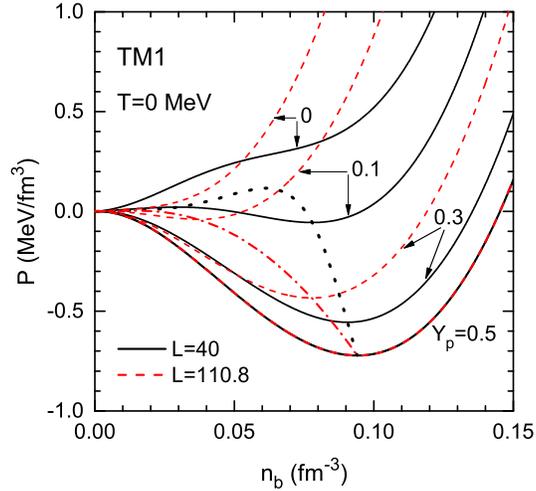}
\caption{(Color online) Pressure of uniform matter $P$ as a
function of baryon density $n_b$ at zero temperature
for various proton fraction $Y_p$.
The black solid and red dashed lines are the results of $L=40$ MeV
and $L=110.8$ MeV in the TM1 set, respectively.
The dotted and dashed-dotted lines indicate
the mechanically unstable regions from negative
compressibility ($dP/dn_b<0$).}
\label{fig:8LPnb}
\end{figure}

The influence of $L$ on the boundary of the liquid-gas coexistence region
is shown in Figs.~\ref{fig:6LTnb} and~\ref{fig:7LYpnb}.
It is seen in Fig.~\ref{fig:6LTnb} that the coexistence region
obtained with a small $L$ is significantly larger than that with a large $L$,
and the maximum $T$ for each $L$ is consistent with the result at $Y_p=0.1$
in Fig.~\ref{fig:5LTcYp}. One can see from Fig.~\ref{fig:7LYpnb} that
the model with a small $L$ predicts a large density range and a small
critical $Y_p$ for the mixed phase. In the case of TM1 at $T=8$ MeV
(see the upper panel of Fig.~\ref{fig:7LYpnb}),
the transition density to uniform matter at $Y_p=0.3$ is about 0.093 fm$^{-3}$
for $L=40$ MeV and 0.087 fm$^{-3}$ for $L=110.8$ MeV.
A comparison of the upper and lower panels indicates that the results
are model dependent, which is also shown clearly in Fig.~11
of Ref.~\cite{Pais15}.
We note from Figs.~\ref{fig:5LTcYp} and~\ref{fig:7LYpnb} that
the $L$ dependence is strongly dependent on $Y_p$,
and there is no difference at $Y_p=0.5$ in one set of models.
The correlation between the symmetry energy slope $L$ and the
boundary of the liquid-gas coexistence region can be understood
from the behavior of the pressure of asymmetric nuclear matter.
It is well known that the pressure of pure neutron matter
is approximately proportional to $L$. In Fig.~\ref{fig:8LPnb},
we show the pressure of uniform matter as a function of
the baryon density $n_b$ at $T=0$ for various $Y_p$ using the models
with $L=40$ MeV and $L=110.8$ MeV in the TM1 set.
It is evident that the model with a small $L$ yields relatively low
pressures, which implies a large coexistence region,
where the dotted and dashed-dotted lines indicate
the mechanically unstable regions from negative
compressibility ($dP/dn_b<0$).

\begin{figure}[htb]
\includegraphics[bb=38 161 564 665, width=7 cm,clip]{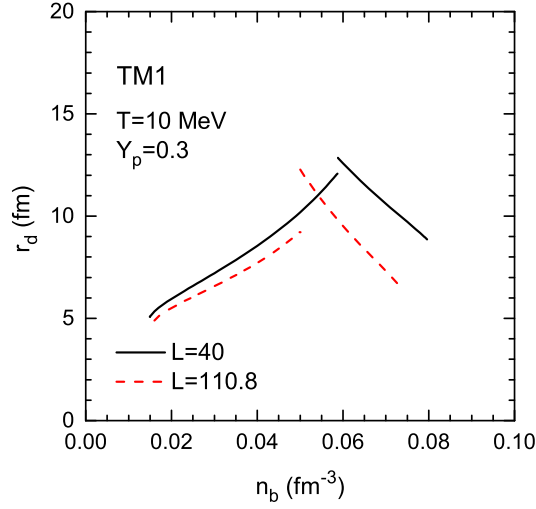}
\caption{(Color online) Radius of the droplet or bubble $r_d$ as a function of
$n_b$ at $T=10$ MeV and $Y_p=0.3$ using the models with
$L=40$ and $110.8$ MeV in the TM1 set.}
\label{fig:9LRnb}
\end{figure}

\begin{figure}[htb]
\includegraphics[bb=24 162 558 664, width=7 cm,clip]{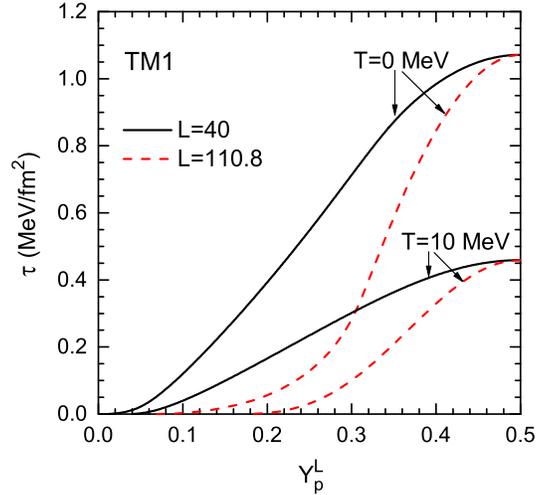}
\caption{(Color online) Surface tension $\tau$ as a function of proton
fraction in the liquid phase $Y^L_p$ at $T=0$ and $10$ MeV
using the models with $L=40$ and $110.8$ MeV in the TM1 set. }
\label{fig:10LTaoYp}
\end{figure}

\begin{center}
\begin{figure}[thb]
\centering
\begin{tabular}{cc}
\includegraphics[bb=18 168 564 665, width=0.45\linewidth, clip]{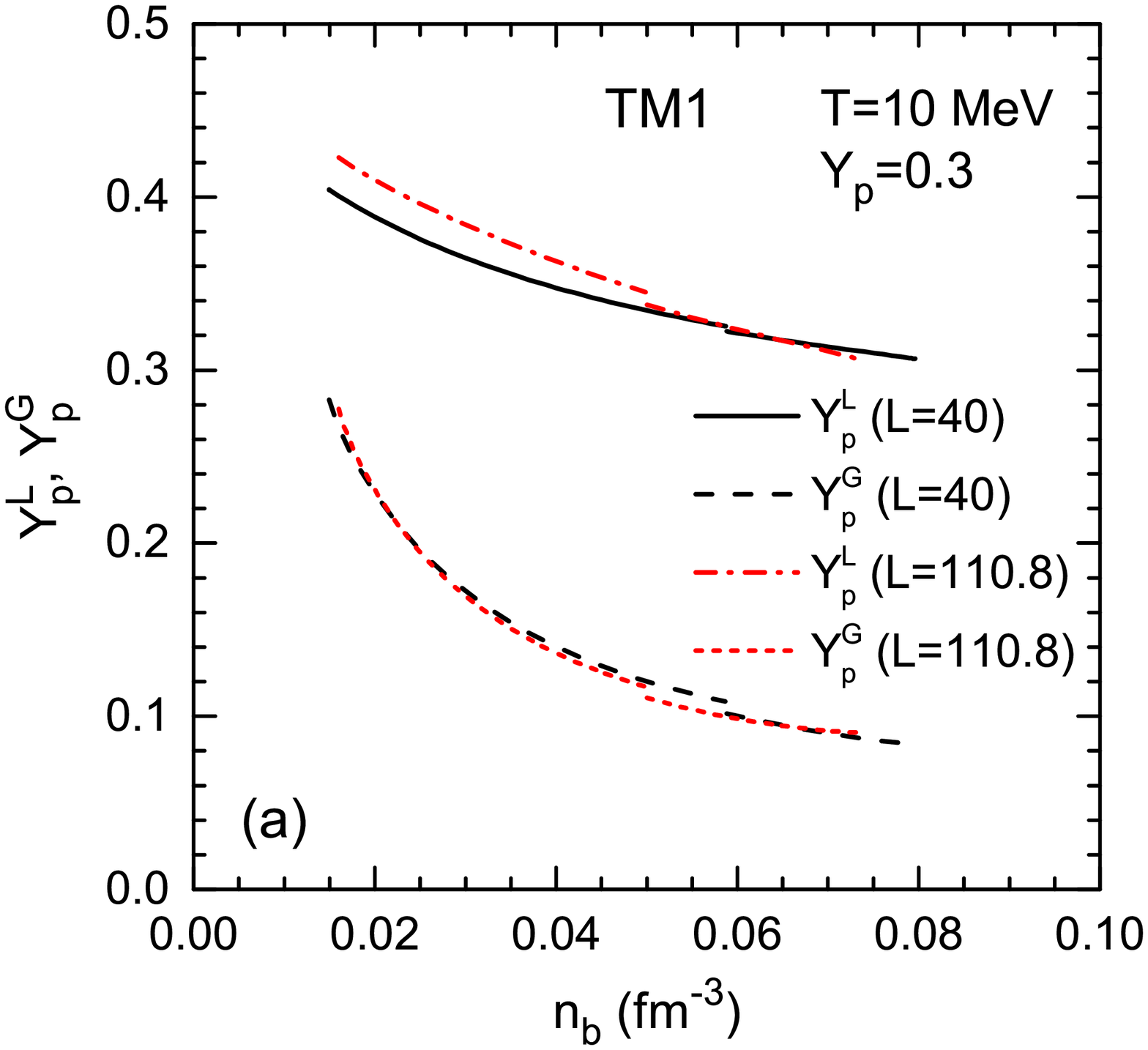}&
\includegraphics[bb=18 168 564 665, width=0.45\linewidth, clip]{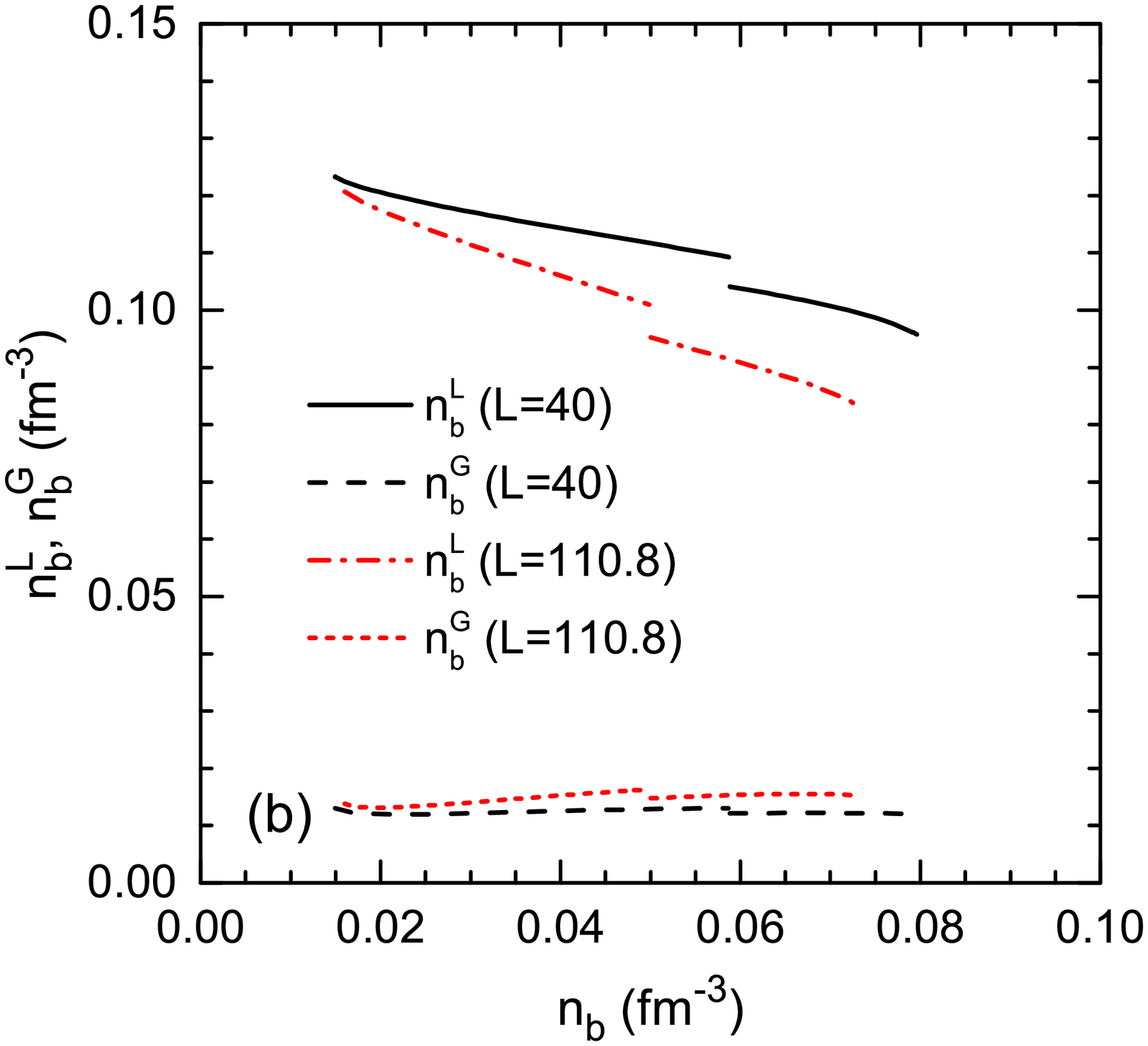}  \\
\includegraphics[bb=18 168 564 665, width=0.45\linewidth, clip]{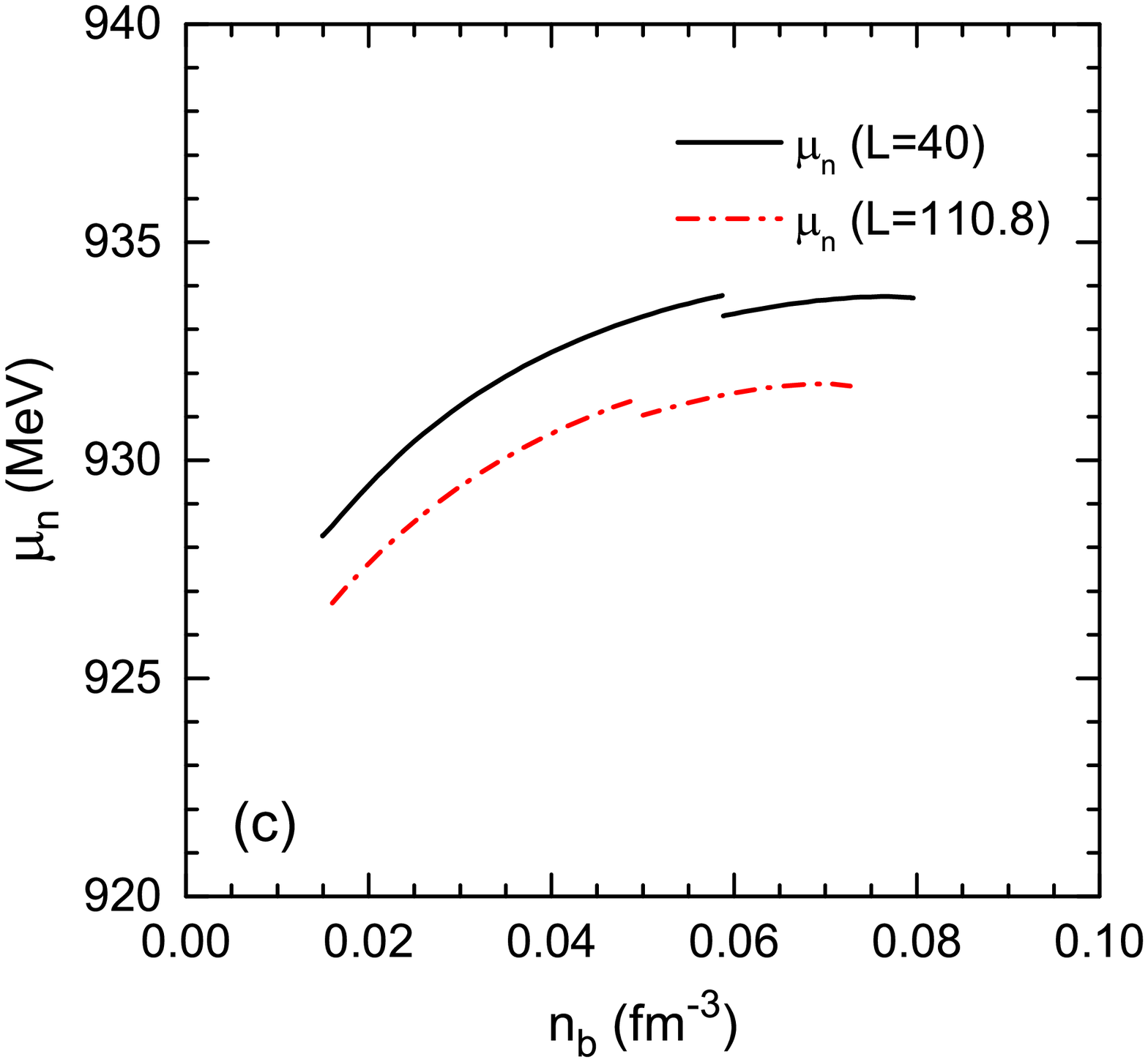}&
\includegraphics[bb=18 168 564 665, width=0.45\linewidth, clip]{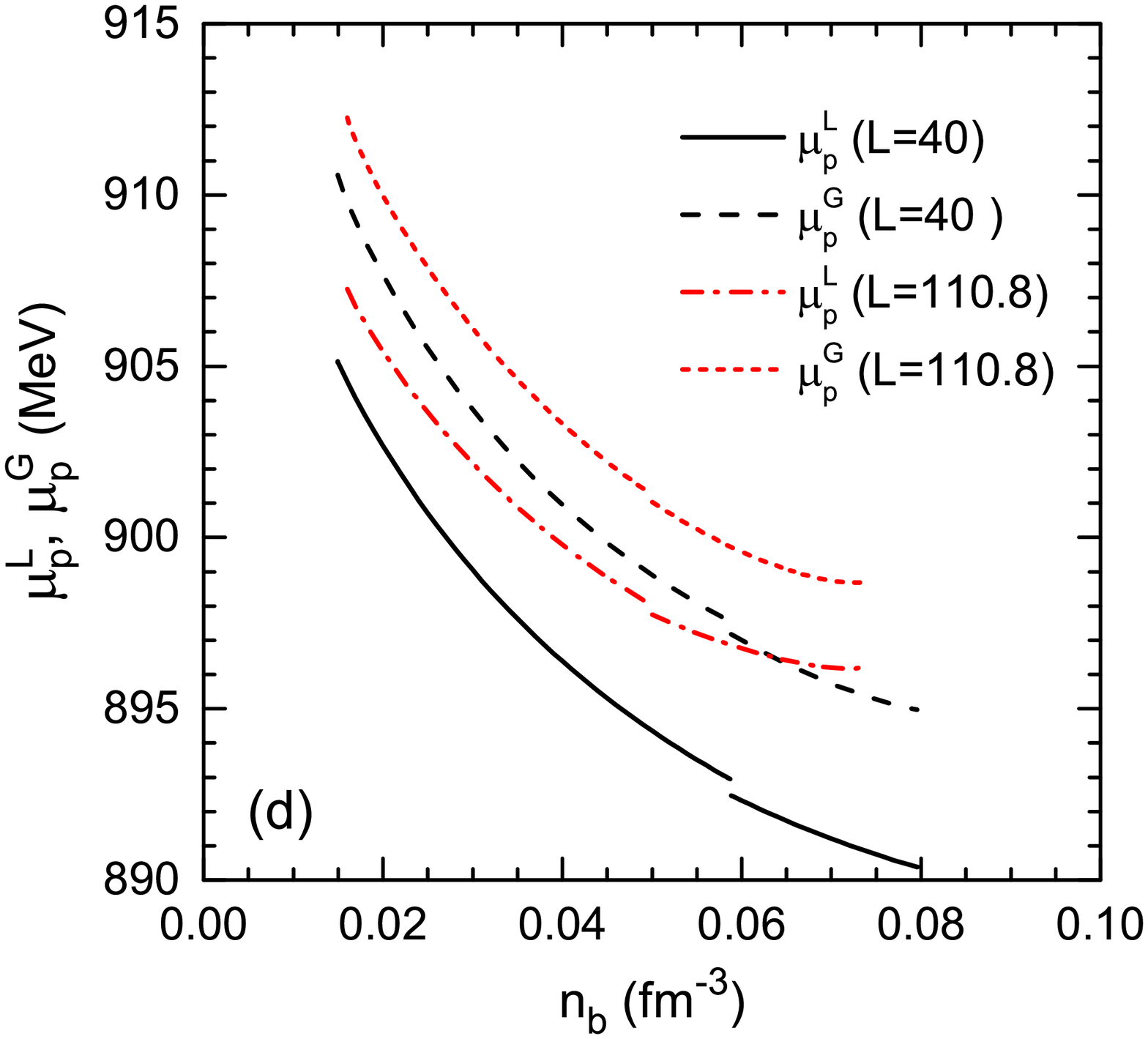} \\
\end{tabular}
\caption{(Color online) Properties of the liquid ($L$) and gas ($G$) mixed
phase at $T=10$ MeV and $Y_p=0.3$ using the models with $L=40$
and $110.8$ MeV in the TM1 set.
Proton fractions $Y^L_p$ and $Y^G_p$ (a),
baryon densities $n^L_b$ and $n^G_b$ (b),
neutron chemical potential $\mu_n$ (c),
and proton chemical potentials $\mu^L_p$ and $\mu^G_p$ (d)
are plotted as a function of the average baryon density $n_b$.}
\label{fig:11LYpnb}
\end{figure}
\end{center}

There are clear correlations between the properties of the liquid-gas
mixed phase and the symmetry energy slope $L$.
In Fig.~\ref{fig:9LRnb}, we plot the radius of the
droplet or bubble, $r_d$, as a function of $n_b$ at $T=10$ MeV and $Y_p=0.3$.
The results are obtained from the models with $L=40$ MeV and $L=110.8$ MeV
in the TM1 set. It is found that as $n_b$ increases, $r_d$ increases in
the droplet phase and decreases in the bubble phase. This behavior is mainly
from the increase of the liquid volume fraction $u$, which can be seen
from Eq.~(\ref{eq:RD}). By comparing the results obtained with different $L$,
we find that a small $L$ corresponds to a large $r_d$ in both droplet
and bubble configurations. This is because a small $L$ favors a large
surface tension $\tau$, and a large $\tau$ would result in a large $r_d$
as indicated in Eq.~(\ref{eq:RD}). The surface tension plays an important
role in determining properties of the mixed phase.
We calculate the surface tension by using a Thomas-Fermi approach for a
one-dimensional nuclear system as described in Refs.~\cite{Bao14a,Mene10,Agra14}.
In Fig.~\ref{fig:10LTaoYp}, we plot the surface tension $\tau$ as a function
of the proton fraction in the liquid phase $Y_p^L$ at $T=0$ and 10 MeV
for $L=40$ and 110.8 MeV in the TM1 set.
It is evident that $\tau$ decreases with increasing $T$ and with increasing $L$.
At a given $T$, the values of $\tau$ for different $L$ are identical at $Y_p^L=0.5$,
which is because the models with different $L$ have the same properties
of symmetric nuclear matter. As $Y_p^L$ decreases, $\tau$ decreases
monotonically and shows a clear dependence on $L$.
We examine the $L$ dependence of properties of the coexisting liquid and
gas phases. We compare results obtained with $L=40$ MeV and $L=110.8$ MeV
in the TM1 set at $T=10$ MeV and $Y_p=0.3$.
In Fig.~\ref{fig:11LYpnb}, we present the following
quantities in the liquid and gas phases:
(a) proton fractions $Y_p^L$ and $Y_p^G$;
(b) baryon densities $n_b^L$ and $n_b^G$;
(c) neutron chemical potential $\mu_n=\mu_n^L=\mu_n^G$;
(d) proton chemical potentials $\mu_p^L$ and $\mu_p^G$.
As one can see from Fig.~\ref{fig:11LYpnb}(a), both $Y_p^L$ and $Y_p^G$
decrease with increasing $n_b$. At low density in the droplet phase,
$Y_p^L$ obtained with $L=110.8$ MeV is somewhat larger than that obtained
with $L=40$ MeV. This is because a large $L$ corresponds to a high
symmetry energy at $n_b>0.11$ fm$^{-3}$, and a high symmetry energy
favors a large proton fraction. On the other hand, the difference of
$Y_p^L$ between $L=40$ MeV and $L=110.8$ MeV is quite small in the bubble
phase. It is seen from Fig.~\ref{fig:11LYpnb}(b) that, with increasing
$n_b$, the baryon density of the liquid phase $n_b^L$ decreases significantly,
and the difference of $n_b^L$ between $L=40$ MeV and $L=110.8$ MeV
becomes larger and larger. Because the model with $L=40$ MeV has relatively
large $g_{\rho}$ and $\Lambda_{\rm{v}}$
(see Table $\textrm{\uppercase\expandafter{\romannumeral2}}$
of Ref.~\cite{Bao14b}),
it would lead to a small value of $g_{\omega}\omega$ and a large negative
value of $g_{\rho}\rho$ in comparison with the case of $L=110.8$ MeV.
To satisfy the equilibrium conditions expressed in
Eqs.~(\ref{eq:cmun})--(\ref{eq:cp}), the model with a small $L$
yields a large $n_b^L$ and a small $n_b^G$, meanwhile,
it results in large neutron chemical potentials, as shown in
Fig.~\ref{fig:11LYpnb}(c), and small proton chemical potentials,
as shown in Fig.~\ref{fig:11LYpnb}(d). Therefore, we conclude that
the properties of the coexisting liquid and gas phases are evidently
dependent on the symmetry energy slope $L$.

\section{Conclusions}
\label{sec:4}

We have investigated the finite-size effect on the liquid-gas phase
transition of stellar matter. The CLD model was used to describe the nucleus embedded
in a gas of electrons and nucleons at finite temperature.
We have employed the Wigner-Seitz approximation to describe the nonuniform
matter in the liquid-gas coexistence region.
The equilibrium conditions for coexisting phases have been
derived by minimization of the total free energy including the surface and
Coulomb contributions. It was found that these equilibrium conditions
are different from the Gibbs conditions used in the bulk calculations
because of the inclusion of surface and Coulomb terms.
We have found that the finite-size effect could significantly reduce
the region of the liquid-gas mixed phase. The critical temperatures
obtained with finite-size effects are much lower than those obtained
from a bulk calculation, and moreover, the maximum critical temperature
with finite-size effects could not be achieved at $Y_p=0.5$, because the
isospin symmetry of the nuclear system is broken by Coulomb interaction.
We have made a detailed comparison of the properties of the liquid-gas
mixed phase with and without finite-size effects. It was found
that there are noticeable differences in properties such as pressures
and chemical potentials.

The influence of the symmetry energy and its slope on the liquid-gas
phase transition of stellar matter was examined with the inclusion
of finite-size effects.
We have employed two sets of generated models based on the TM1
and IUFSU parametrizations, where all models in each set have the same
isoscalar saturation properties and fixed symmetry energy at a density
of $0.11\, \rm{fm}^{-3}$ but have different symmetry energy slope $L$.
By using these models, we have found that there are considerable
differences in the critical temperature $T_c$ at low $Y_p$ region
between the models with different $L$. The model with a small $L$
predicts a high $T_c$. The boundary of the liquid-gas coexistence region
was found to be related to the symmetry energy slope $L$.
At a fixed temperature, the model with a small $L$ predicts a large
density range and a small critical $Y_p$ for the mixed phase.
The surface tension plays an important role in determining properties
of the coexisting liquid and gas phases. It was found that
a small $L$ corresponds to a large surface tension $\tau$, which results
in a large radius of the droplet or bubble.
We note that only droplet and bubble configurations have been considered
in the present work. It would be interesting to include other pasta phases,
such as rod, slab, and tube, which may appear in the middle density region
and can smooth the transition from droplet to bubble.

\section*{Acknowledgment}

This work was supported in part by the National Natural
Science Foundation of China (Grant No. 11375089).


\newpage
\begin{table}[tbp]
\caption{Parameter sets used in this work. The masses are given in MeV.}
\begin{center}
\begin{tabular}{lccccccccccc}
\hline\hline
Model   &$M$  &$m_{\sigma}$  &$m_\omega$  &$m_\rho$  &$g_\sigma$  &$g_\omega$
        &$g_\rho$ &$g_{2}$ (fm$^{-1}$) &$g_{3}$ &$c_{3}$ &$\Lambda_{\textrm{v}}$ \\
\hline
TM1     &938.0  &511.198  &783.0  &770.0  &10.0289  &12.6139  &9.2644
        &$-$7.2325   &0.6183   &71.3075   &0.000  \\
IUFSU   &939.0  &491.500  &782.5  &763.0  &9.9713   &13.0321  &13.5900
        &$-$8.4929   &0.4877   &144.2195  &0.046 \\
\hline\hline
\end{tabular}
\label{tab:1}
\end{center}
\end{table}

\begin{table}[htb]
\caption{Saturation properties of nuclear matter for the TM1 and IUFSU models.
The quantities $E_0$, $K$, $E_{\text{sym}}$, and $L$ are, respectively,
the energy per nucleon, incompressibility coefficient, symmetry
energy, and symmetry energy slope at saturation density $n_0$.}
\label{tab:2}
\begin{center}
\begin{tabular}{lccccc}
\hline\hline
Model & $n_0$ (fm$^{-3}$) & $E_0$ (MeV) & $K$ (MeV) & $E_{\text{sym}}$ (MeV) & $L$ (MeV) \\
\hline
TM1   & 0.145 & $-$16.3 & 281 & 36.9 & 110.8 \\
IUFSU & 0.155 & $-$16.4 & 231 & 31.3 & 47.2  \\
\hline\hline
\end{tabular}
\end{center}
\end{table}

\newpage

\begin{thebibliography}{99}

\bibitem{PR05} C. B. Das, S. Das Gupta, W. G. Lynch, A. Z. Mekjian, and
M. B. Tsang, Phys. Rep. \textbf{406}, 1 (2005).

\bibitem{LiBA08} B. A. Li, L. W. Chen, and C. M. Ko,
Phys. Rep. \textbf{464}, 113 (2008).

\bibitem{Pal10b} B. K. Sharma and S. Pal,
Phys. Rev. C \textbf{82}, 055802 (2010).

\bibitem{Hemp10} M. Hempel and J. Schaffner-Bielich,
Nucl. Phys. A \textbf{837}, 210 (2010).

\bibitem{Sagu14} V. V. Sagun, A. I. Ivanytskyi, K. A. Bugaev, and I. N. Mishustin,
Nucl. Phys. A \textbf{924}, 24 (2014).

\bibitem{Jaqa84} H. Jaqaman, A. Z. Mekjian, and L. Zamick,
Phys. Rev. C \textbf{27}, 2782 (1983); \textbf{29}, 2067 (1984).

\bibitem{Latt85} J. M. Lattimer, C. J. Pethick, D. G. Ravenhall, and D. Q. Lamb,
Nucl. Phys. A \textbf{432}, 646 (1985).

\bibitem{Sero95} H. M\"{u}ller and B. D. Serot,
Phys. Rev. C \textbf{52}, 2072 (1995).

\bibitem{PRL95} J. Pochodzalla \textit{et al}.,
Phys. Rev. Lett. \textbf{75}, 1040 (1995).

\bibitem{PR04} Ph. Chomaz, M. Colonna, and J. Randrup,
Phys. Rep. \textbf{389}, 263 (2004).

\bibitem{RPP05} S. Shlomo and V. M. Kolomietz,
Rep. Prog. Phys. \textbf{68}, 1 (2005).

\bibitem{PPNP08} B. Borderie and M. F. Rivet,
Prog. Part. Nucl. Phys. \textbf{61}, 551 (2008).

\bibitem{Pal10a} B. K. Sharma and S. Pal,
Phys. Rev. C \textbf{81}, 064304 (2010).

\bibitem{Marg03} J. Margueron and Ph. Chomaz,
Phys. Rev. C \textbf{67}, 041602(R) (2003).

\bibitem{Bara05} V. Baran, M. Colonna, V. Greco, and M. Di Toro,
Phys. Rep. \textbf{410}, 335 (2005).

\bibitem{SSA06} S. S. Avancini, L. Brito, Ph. Chomaz, D. P. Menezes,
and C. Provid\^{e}ncia, Phys. Rev. C \textbf{74}, 024317 (2006).

\bibitem{Duco06} C. Ducoin, Ph. Chomaz, and F. Gulminelli,
Nucl. Phys. A \textbf{771}, 68 (2006).

\bibitem{Duco07} C. Ducoin, K. H. O. Hasnaoui, P. Napolitani, Ph. Chomaz,
and F. Gulminelli, Phys. Rev. C \textbf{75}, 065805 (2007).

\bibitem{Hemp13} M. Hempel, V. Dexheimer, S. Schramm, and I. Iosilevskiy,
Phys. Rev. C \textbf{88}, 041906 (2013).

\bibitem{Well14} C. Wellenhofer, J. W. Holt, N. Kaiser, and W. Weise,
Phys. Rev. C \textbf{89}, 064009 (2014).

\bibitem{Well15} C. Wellenhofer, J. W. Holt, and N. Kaiser,
Phys. Rev. C \textbf{92}, 015801 (2015).

\bibitem{Lee01} S. J. Lee and A. Z. Mekjian,
Phys. Rev. C \textbf{63}, 044605 (2001).

\bibitem{Pawl02} P. Pawlowski, Phys. Rev. C \textbf{65}, 044615 (2002).

\bibitem{Sil04} T. Sil, S. K. Samaddar, J. N. De, and S. Shlomo,
Phys. Rev. C \textbf{69} 014602 (2004).

\bibitem{Maru10} T. Maruyama, N. Yasutake, and T. Tatsumi,
Prog. Theor. Phys. Suppl. \textbf{186}, 69 (2010).

\bibitem{Bao14b} S. S. Bao, J. N. Hu, Z. W. Zhang, and H. Shen,
Phys. Rev. C \textbf{90}, 045802 (2014).

\bibitem{Mene08} S. S. Avancini, D. P. Menezes, M. D. Alloy, J. R.
Marinelli, M. M. W. Moraes, and C. Provid\^{e}ncia,
Phys. Rev. C \textbf{78}, 015802 (2008).

\bibitem{Bao14a} S. S. Bao and H. Shen,
Phys. Rev. C \textbf{89}, 045807 (2014).

\bibitem{Pais15} H. Pais, S. Chiacchiera, and C. Provid\^{e}ncia,
Phys. Rev. C \textbf{91}, 055801 (2015).

\bibitem{Maru05} T. Maruyama, T. Tatsumi, D. N. Voskresensky, T. Tanigawa,
and S. Chiba, Phys. Rev. C \textbf{72}, 015802 (2005).

\bibitem{Mene10} S. S. Avancini, S. Chiacchiera, D. P. Menezes,
and C. Provid\^{e}ncia, Phys. Rev. C \textbf{82}, 055807 (2010);
\textbf{85}, 059904(E) (2012).

\bibitem{Agra14} B. K. Agrawal, D. Bandyopadhyay, J. N. De, and S. K. Samaddar,
Phys. Rev. C \textbf{89} 044320 (2014).

\bibitem{Cent98} M. Centelles, M. Del Estal, and X. Vi\~{n}as,
Nucl. Phys. A \textbf{635}, 193 (1998).

\bibitem{Douc00} F. Douchin, P. Haensel, and J. Meyer,
Nucl. Phys. A \textbf{665}, 419 (2000).

\bibitem{Wata04} G. Watanabe, K. Sato, K. Yasuoka, and T. Ebisuzaki,
Phys. Rev. C \textbf{69}, 055805 (2004).

\bibitem{Wata08} H. Sonoda, G. Watanabe, K. Sato, K. Yasuoka, and T. Ebisuzaki,
Phys. Rev. C \textbf{77}, 035806 (2008).

\bibitem{Oyam07} K. Oyamatsu and K. Iida,
Phys. Rev. C \textbf{75}, 015801 (2007).

\bibitem{Hemp12} M. Hempel, T. Fischer, J. Schaffner-Bielich, and M. Liebend\"{o}rfer,
Astrophys. J. \textbf{748}, 70 (2012).

\bibitem{Radu10} A. R. Raduta and F. Gulminelli,
Phys. Rev. C \textbf{82}, 065801 (2010).

\bibitem{Gulm15} F. Gulminelli and A. R. Raduta,
Phys. Rev. C \textbf{92}, 055803 (2015).

\bibitem{Furu11} S. Furusawa, S. Yamada, K. Sumiyoshi, and H. Suzuki,
Astrophys. J. \textbf{738}, 178 (2011).

\bibitem{Furu13} S. Furusawa, K. Sumiyoshi, S. Yamada, and H. Suzuki,
Astrophys. J. \textbf{772}, 95 (2013).

\bibitem{Botv13} N. Buyukcizmeci, A. S. Botvina, I. N. Mishustin, R. Ogul,
M. Hempel, J. Schaffner-Bielich, F.-K. Thielemann, S. Furusawa, K. Sumiyoshi,
S. Yamada, and H. Suzuki,
Nucl. Phys. A \textbf{907}, 13 (2013).

\bibitem{Duco10} C. Ducoin, J. Margueron, and C. Provid\^{e}ncia,
Europhys. Lett. \textbf{91}, 32001 (2010).

\bibitem{Chen13} Z. Zhang and L. W. Chen,
Phys. Lett. B \textbf{726}, 234 (2013).

\bibitem{Chen07} J. Xu, L. W. Chen, B. A. Li, and H. R. Ma,
Phys. Lett. B \textbf{650}, 348 (2007).

\bibitem{Jiang13} G. H. Zhang and W. Z. Jiang,
Phys. Lett. B \textbf{720}, 148 (2013).

\bibitem{Sero97} B. D. Serot and J. D. Walecka,
Int. J. Mod. Phys. E \textbf{06}, 515 (1997).

\bibitem{Bogu77} J. Boguta and A. R. Bodmer,
Nucl. Phys. A \textbf{292}, 413 (1977).

\bibitem{Ring90} Y. K. Gambhir, P. Ring, and A. Thimet, Ann. Phys. (N.Y.)
\textbf{198}, 132 (1990).

\bibitem{Meng06} J. Meng, H. Toki, S. G. Zhou, S. Q. Zhang, W. H. Long, and
L. S. Geng, Prog. Part. Nucl. Phys. \textbf{57}, 470 (2006).

\bibitem{TM1} Y. Sugahara and H. Toki, Nucl. Phys. A \textbf{579}, 557 (1994).

\bibitem{IUFSU} F. J. Fattoyev, C. J. Horowitz, J. Piekarewicz, and G. Shen,
Phys. Rev. C \textbf{82}, 055803 (2010).

\bibitem{Shen02} H. Shen, Phys. Rev. C \textbf{65}, 035802 (2002).

\bibitem{Shen11} H. Shen, H. Toki, K. Oyamatsu, and K. Sumiyoshi, Astrophys. J.
Suppl. \textbf{197}, 20 (2011).

\bibitem{Zhang14} Z. W. Zhang and H. Shen,
Astrophys. J. \textbf{788}, 185 (2014).

\bibitem{FSU} B. G. Todd-Rutel and J. Piekarewicz,
Phys. Rev. Lett. \textbf{95}, 122501 (2005).

\bibitem{Horo01} C. J. Horowitz and J. Piekarewicz,
Phys. Rev. Lett. \textbf{86}, 5647 (2001).

\bibitem{Horo03} J. Carriere, C. J. Horowitz, and J. Piekarewicz,
Astrophys. J. \textbf{593}, 463 (2003).

\bibitem{Mene11} R. Cavagnoli, D. P. Menezes, and C. Provid\^{e}ncia,
Phys. Rev. C \textbf{84}, 065810 (2011).

\bibitem{Prov13} C. Provid\^{e}ncia and A. Rabhi,
Phys. Rev. C \textbf{87}, 055801 (2013).

\bibitem{Latt91} J. M. Lattimer and F. D. Swesty,
Nucl. Phys. A \textbf{535}, 331 (1991).

\bibitem{Wata00} G. Watanabe, K. Iida, and K. Sato,
Nucl. Phys. A \textbf{676}, 455 (2000);
\textbf{726}, 357 (2003).

\bibitem{Chamel08} N. Chamel and P. Haensel,
Living Rev. Relativ. \textbf{11}, 10 (2008).

\bibitem{Bao15} S. S. Bao and H. Shen,
Phys. Rev. C \textbf{91}, 015807 (2015).

\end{thebibliography}
\end{document}